\documentstyle{mn}

\title[Quasar host galaxy spectra]{Optical off-nuclear spectra of quasar hosts and radio galaxies}

\author[D. H. Hughes et al.]
{David~H.~Hughes$^{1}$\thanks{current address: INAOE, Apartado Postal 51 y 216, 7200, Puebla, Pue., Mexico}, Marek~J.~Kukula$^{1}$, James~S.~Dunlop$^{1}$ and Todd Boroson$^{2}$\\
$^{1}$~Institute for Astronomy, Department of Physics and Astronomy,
University of Edinburgh, Blackford Hill, Edinburgh EH9 3HJ, U.K. \\
$^{2}$~NOAO, PO Box 26732, Tucson, Arizona 85726-6732, U.S.A. \\}

\date{}

\begin{document}
\maketitle

\begin{abstract}

We present optical ($\sim 3200$\AA~ to $\sim9000$\AA~) off-nuclear
spectra of 26 powerful active galaxies in the redshift range $0.1 \leq
z \leq 0.3$, obtained with the Mayall and William Herschel 4-meter
class telescopes. The sample consists of radio-quiet quasars,
radio-loud quasars (all with $-23 \geq M_{V} \geq -26$) and radio
galaxies of Fanaroff \& Riley Type II (with extended radio
luminosities and spectral indices comparable to those of the
radio-loud quasars).  The spectra were all taken approximately 5
arcseconds off-nucleus, with offsets carefully selected so as to
maximise the amount of galaxy light falling into the slit, whilst
simultaneously minimising the amount of scattered nuclear light.  The
majority of the resulting spectra appear to be dominated by the
integrated stellar continuum of the underlying galaxies rather than by
light from the non-stellar processes occurring in the active nuclei,
and in many cases a 4000\AA~break feature can be identified. The
individual spectra are described in detail, and the importance of the
various spectral components is discussed.  Stellar population
synthesis modelling of the spectra will follow in a subsequent paper
(Nolan et al. 2000).

\end{abstract}

\begin{keywords}
galaxies: stellar content -- galaxies: active -- quasars: general
\end{keywords}

\section{Introduction}

Understanding the host galaxies of active galactic nuclei (AGN) is now
recognised as an important step on the path towards reaching an
understanding of AGN themselves - how they form, how they are fuelled
and how the differences between the various classes of object
arise. In two areas in particular the nature of the host galaxy gives
us a direct insight into the workings of the AGN: galaxy properties
seem to play a role in determining the radio loudness of the central
engine (powerful radio sources are almost never found in spiral or
disc-dominated systems -- though see McHardy et al. 1994); and amongst
radio-loud objects the host galaxies offer a powerful,
orientation-independent means of testing models which attempt to unify
different types of AGN via beaming and viewing angle effects ({\it eg}
Urry \& Padovani 1995). The observations presented here attempt to
address both of these issues by comparing the galaxies associated with
the three main types of powerful AGN: radio-quiet quasars (RQQs),
radio-loud quasars (RLQs) and radio galaxies (RGs) of Fanaroff \&
Riley Type II.

\subsection{Quasar hosts}

Only once the problem of separating the diffuse galaxy emission from
the wings of the quasar point spread function (PSF) has been overcome
can one can begin to describe and classify the morphologies,
brightness profiles and interaction histories of the quasar hosts.
Over the last decade improvements in ground-based techniques and the
advent of the Hubble Space Telescope (HST) have revolutionised our
understanding of quasar host galaxies.

Evidence for mergers or interactions in the form of morphological
disturbances and close companions is a common feature of these images,
but a significant number of quasars are also found in what appear to
be undisturbed hosts. In addition, the idea that radio-loudness is a
straightforward consequence of the host galaxy type has had to be
abandoned.  Although some radio-quiet quasars are found in spirals
({\it eg} Hutchings et al. 1994; \"{O}rndahl, R\"{o}nnback \& van
Groningen 1997) in general the hosts of both radio-loud and
radio-quiet quasars tend to have properties consistent with early-type
galaxies ({\it eg} V\'{e}ron-Cetty \& Woltjer 1990; Disney et
al. 1995, Hutchings \& Morris 1995, Bahcall et al. 1997, McLeod, Rieke
\& Storrie-Lombardi 1999) and typically have luminosities $> L^{*}$
({\it eg} Dunlop et al. 1993, Bahcall, Kirhakos \& Schneider 1994,
1995ab, 1996; Hutchings et al. 1994, Boyce et al.  1998, Hooper, Impey
\& Foltz 1997) and in many cases are comparable in mass to brightest
cluster galaxies (BCGs).

Meanwhile, McLeod \& Rieke (1995a) find evidence for a {\it lower}
limit on the $H$-band luminosities (and hence the mass of the red,
established stellar populations) of galaxies hosting radio-quiet AGN,
which appears to increase as the nuclear luminosity increases,
implying that the nuclear activity is closely linked to the mass of
the bulge component of the host. This impression has been reinforced
by McLure et al. (1999), who find that {\it all} the RQQs with
$M_{R}\leq -24$ in their HST sample occur in massive elliptical
galaxies, with only the least luminous radio-quiet objects lying in
disc-dominated hosts.

Many long-held views about the triggering of nuclear activity and the
origins of radio loudness are currently being reassessed in the light
of these imaging studies, but images cannot tell the whole story. A
completely independent way of characterising the host galaxies of AGN
is via analysis and classification of their stellar populations and
starformation histories. The aim of the observations described in this
paper was to obtain high signal-to-noise spectra of quasar hosts and
radio galaxies for use in spectrophotometric modelling to determine
the nature and history of their stelllar constituents.

\subsection{Spectroscopy of quasar hosts}
  
Previous off-nuclear spectroscopy of the host galaxies of quasars has
produced mixed results. Boroson \& Oke (1982) were the first to detect
an unequivocably stellar continuum from the nebulosity surrounding the
radio-loud quasar 3C48 and subsequent studies revealed stellar
continua and emission/absorption features around several other quasars
(Boroson, Oke \& Green 1982; Boroson \& Oke 1984; Boroson, Persson \&
Oke 1985; Hickson \& Hutchings 1987; Hutchings \& Crampton
1990). However, except for the general result that there appeared to
be systematic differences between the spectra of radio-loud and
radio-quiet quasar hosts, these studies produced little advance in our
understanding of the relationship between RQQs, RLQs and RGs for the
reasons outlined below.

In general the quasar targets were chosen virtually at random,
often for the sole reason that they were `interesting' and/or
unusual. Until now a programme of off-nuclear spectroscopy for
statistically useful and properly matched samples of RQQs and RLQs has
never been carried out, nor has any attempt been made to compare the
off-nuclear spectra of RLQs with equivalent {\it off-nuclear} spectra
of radio galaxies. 

The early work also focussed mostly on emission-line activity, with
discussion of the stellar continuum being confined to classification
as red or blue, and the identification of a few stellar features. Only
limited attempts were made to use the form of the spectrum to
investigate the composition and evolution of the stellar population.

Finally, few of the host galaxy spectra were taken sufficiently
off-nucleus - {\it eg} the spectra taken by Boroson and collaborators
were taken only 3$''$ from the quasar. This was done in the belief
that any further off-nucleus the host galaxy would be too faint for a
reasonable spectrum to be obtained, but inevitably resulted in
significant contamination of the host galaxy spectrum by scattered
light from the quasar nucleus. A scaled version of the quasar spectrum
had therefore to be subtracted from the off-nuclear spectrum to reveal
the spectrum of the underlying host, but the extra noise and
systematic errors introduced by this process severely limited the
quality of the final spectra.

Thus, the main obstacles in these previous attempts to classify the
hosts of powerful AGN were the difficulty in separating the underlying
starlight from the glare of the quasar and - less immediately, but
still of some importance - the lack of well-defined samples of a
sufficient size to carry out straightforward statistical comparisons.

As far as possible, the design of the current study was intended to
circumvent both these problems with the goal of obtaining clean
spectra of the stellar component of radio-loud and radio-quiet quasar
hosts and radio galaxies, which could then be used to search for
systematic differences and similarities between the stellar
populations and of the galaxies hosting each type of activity. Despite
that fact that the radio galaxies lack bright nuclear point sources we
were careful to adopt exactly the same observing strategy as used for
the quasars to ensure that the data would be directly comparable.

In this paper we describe the observations and present the spectra.  A
second paper (Nolan et al. 2000) will describe the results of
spectrophotometric modelling to estimate the ages and starformation
histories of the galaxies (preliminary results have already been
reported by Kukula et al. 1997).  The current paper is organised as
follows. Section~2 describes the samples used and Section~3 details
the observing and data reduction strategies chosen to optimise the
amount of starlight collected by the various instruments employed
throughout the study. Section~4 gives an overview of the data obtained
and Sections~5 and 6 contain more detailed information on the
individual spectra.

\section{Sample selection}

Dunlop et al. (1993) and Taylor et al. (1996) observed a sample of
intermediate-redshift ($0.1 \leq z\leq 0.3$) radio-loud and -quiet
quasars\footnote{Dunlop et al. define `radio quiet' objects as those with
$L_{5GHz} < 10^{24}$W~Hz$^{-1}$sr$^{-1}$.} and Fanaroff-Riley Type II
radio galaxies (RGs) in the near infrared ($K$-band:2.2$\mu$m) in
order to compare the luminosities and morphologies of the galaxies
associated with these three main types of powerful active nucleus. The
choice of waveband was informed by the low quasar:host ratio in the
near infrared, which allowed accurate determination of the galaxy
properties from the ground with the minimum amount of confusion from
the point spread function of the central quasar. Their sample was
carefully constructed to ensure that the different types of object
could be compared directly with one another: the radio-loud and -quiet
quasars both have the same distribution of optical luminosities ($-23
\geq M_{V} \geq -26$) and both the radio-loud quasars and the radio
galaxies have similar extended radio luminosities and morphologies and
steep radio spectra.  The sample is therefore ideal for investigating
the influence of galaxy properties on the `radio loudness' of
otherwise very similar quasars, and also for testing unified models of
RLQs and RGs which predict that the properties of their hosts should
be identical.  A substantial subset of the sample has also formed the
basis for an $R$-band imaging study using HST (Kukula et al. 1999,
McLure et al. 1999, Dunlop et al. 2000), in which the enhanced angular
resolution of HST has allowed both unambiguous identification of the
host morphology and identification of detailed substructure in the
quasar hosts and radio galaxies.

This sample provides an ideal starting point for a spectroscopic
study of host galaxies because, in addition to the careful selection
criteria, the existence of deep near-infrared images of every object
provides us with a unique opportunity to minimise the contamination of
the galaxy spectra by quasar light.  Armed with knowledge of the
extent and orientation of the host galaxy on the sky one is able to
choose a slit position which is far enough from the nucleus to avoid
the worst excesses of scattered quasar light, but which simultaneously
maximises the amount of galaxy light falling onto the slit.

Out of the the 40 objects in the original Taylor et al. sample a total
of 26 objects were observed in the current study (9 RQQs, 10 RLQs and
7 RGs). Details are listed in Table~1. Note that the radio source 3C59
has been shown by Meurs \& Unger (1991) to consist of three separate
sources, of which only the weakest appears to be associated with the
quasar 0204+292. As a result, the radio luminosity of 0204+292 places
it below our dividing line of $L_{5GHz} = 10^{24}$W~Hz$^{-1}$sr$^{-1}$
and we classify it as an RQQ.

\begin{table*}
\caption{Objects discussed in the current paper. Column 6 lists the
telescope with which the object's spectrum was obtained: M4M denotes
the Mayall 4-m Telescope at Kitt Peak; WHT denotes the 4.2-m William
Herschel Telescope on La Palma.}

\begin{tabular}{cllcccl}

\hline
\footnotesize
IAU  & \multicolumn{2}{c}{Optical position (J2000)}& V & $z$ & Telescope & Alternative \\
name &  RA ($h~m~s$)  & Dec ($^{\circ}~'~''$)      &   &     &           & names             \\ \hline
\multicolumn{7}{c}{\it Radio-Quiet Quasars} \\ \hline
0007$+$106  &  00 10 30.98 &$+$10 58 28.4  &    15.40 &  0.089 & M4M    &   III Zw 2    \\                     
0054$+$144  &  00 57 09.92 &$+$14 46 11.0  &    15.71 &  0.171 & M4M/WHT&   PHL909      \\                     
0157$+$001  &  01 59 49.72 &$+$00 23 41.0  &    15.69 &  0.164 & M4M/WHT&   Mrk 1014    \\        
0204$+$292  &  02 07 02.16 &$+$29 30 45.4  &    16.00 &  0.109 & WHT    &   3C59        \\                     
0244$+$194  &  02 47 41.22 &$+$19 40 53.6  &    16.66 &  0.176 & WHT    &   MS 02448+19 \\
1549$+$203  &  15 52 02.28 &$+$20 14 01.6  &    16.50 &  0.250 & WHT    &   1E15498+203, LB906, MS 15498+20\\             
1635$+$119  &  16 37 46.54 &$+$11 49 49.8  &    16.50 &  0.146 & WHT    &   MC1635+119,  MC 2   \\             
2215$-$037  &  22 17 47.35 &$-$03 32 48.4  &    17.20 &  0.241 & WHT    &   MS 22152-03, EX2215-037\\
2344$+$184  &  23 47 25.71 &$+$18 44 58.4  &    15.90 &  0.138 & M4M/WHT&   E2344+184\\ \hline                  
\multicolumn{7}{c}{\it Radio-Loud Quasars} \\ \hline
0137$+$012  &  01 39 57.24 &$+$01 31 46.8  &    17.07 &  0.258 & M4M    &   PHL1093       \\                   
0736$+$017  &  07 39 18.01 &$+$01 37 04.6  &    16.47 &  0.191 & M4M/WHT&   S0736+01, OI061 \\              
1004$+$130  &  10 07 26.12 &$+$12 48 56.4  &    15.15 &  0.240 & WHT    &   S1004+13, OL107.7, 4C13.41\\    
1020$-$103  &  10 22 32.79 &$-$10 37 44.2  &    16.11 &  0.197 & M4M    &   S1020-103, OL133 \\              
1217$+$023  &  12 20 11.90 &$+$02 03 42.3  &    16.53 &  0.240 & WHT    &   S1217+02, UM492 \\              
2135$-$147  &  21 37 45.24 &$-$14 32 55.6  &    15.53 &  0.200 & WHT    &   S2135-14, PHL1657\\             
2141$+$175  &  21 43 35.56 &$+$17 43 49.1  &    15.73 &  0.213 & WHT    &   OX169   \\
2201$+$315  &  22 03 15.00 &$+$31 45 38.3  &    15.58 &  0.298 & M4M    &   B2 2201+31A, 4C31.63\\             
2247$+$140  &  22 50 25.40 &$+$14 19 49.9  &    15.33 &  0.237 & M4M/WHT&   PKS2247+14, 4C14.82\\             
2349$-$014  &  23 51 56.14 &$-$01 09 12.8  &    15.33 &  0.173 & WHT    &   PKS2349-01, PB5564\\ \hline              
\multicolumn{7}{c}{\it Radio Galaxies} \\ \hline
0230$-$027  &  02 32 43.14 &$-$02 33 34.1  &    19.20 &  0.239 & WHT    &   PKS0230-027    \\                  
0345$+$337  &  03 48 46.9  &$+$33 53 16    &    19.00 &  0.244 & WHT    &   3C93.1  \\                         
0917$+$459  &  09 21 08.95 &$+$45 38 55.8  &    17.22 &  0.174 & WHT    &   3C219.0 \\            
1215$-$033  &  12 17 55.3  &$-$03 37 23    &    18.90 &  0.184 & WHT    &           \\
1330$+$022  &  13 32 53.27 &$+$02 00 45.0  &    18.27 &  0.215 & M4M    &   3C287.1 \\            
1334$+$008  &  13 37 31.38 &$+$00 35 29.0  &    19.00 &  0.299 & WHT    &           \\                   
2141$+$279  &  21 44 11.66 &$+$28 10 18.9  &    18.15 &  0.215 & M4M/WHT&   3C436   \\ \hline

\end{tabular}
\end{table*}

\begin{table*}
\small
\caption{Objects observed on the Mayall 4-m Telescope at Kitt
Peak. The detector used was a Tek T2KB chip. Columns are as follows:
(1) source name, listed in RA order; (2) type of spectrum, `G' $=$
off-nuclear (galaxy) spectrum, `N' $=$ nuclear spectrum; (3) redshift;
(4) date of observation (dd/mm/yy); (5) average airmass during
observation; (6) average seeing during observation (arcsec); (7)
number and length (in seconds) of consecutive exposures; (8) slit
width (set to $3''$ throughout); (9) wavelength range determined by
grating setting; (10) positional offset of slit centre from
quasar/galaxy centroid, and PA of slit (measured East from North).}

\begin{tabular}{lcccccrccl}

\hline
\footnotesize
Source  & Type & z &Date&Airmass&Seeing&Exposure&Slit &$\lambda$-range&\multicolumn{1}{c}{Slit offset}\\
        &      &   &    &       &($''$)& (sec)  &width&(\AA)          &\multicolumn{1}{c}{\& PA}    \\ \hline
\multicolumn{10}{c}{\it Radio-Quiet Quasars} \\ \hline
0007+106  & G & 0.089&  25/09/92 & 1.10 & 1.7 & 5$\times$1800& 3$''$&3665-7121 &5$''$N, 90$^{\circ}$                \\  
          & N & 0.089&  25/09/92 & 1.13 & 1.6 & 1$\times$1800& 3$''$&3665-7121 &zero offset              \\ 
0054+144  & G & 0.171&  26/09/92 & 1.19 & 1.8 & 6$\times$1800& 3$''$&3659-7121 &3$''$S, 4$''$E, 15$^{\circ}$         \\
          & N & 0.171&  26/09/92 & 1.60 & 1.6 &  1$\times$900& 3$''$&3659-7121 &zero offset              \\
0157+001  & G & 0.164&  25/09/92 & 1.26 & 1.8 & 5$\times$1800& 3$''$&3662-7121 &5$''$N, 90$^{\circ}$		   \\
          & N & 0.164&  25/09/92 & 1.55 & 1.9 & $1\times1200$& 3$''$&3662-7121 &zero offset             \\
2344+184  & G & 0.138&  24/09/92 & 1.05 & 1.9 & 5$\times$1800& 3$''$&3665-7124 &5$''$E, 177$^{\circ}$             	 \\
          & N & 0.138&  24/09/92 & 1.15 & 2.5 & 1$\times$1800& 3$''$&3662-7121 &zero offset       \\ \hline 
\multicolumn{10}{c}{\it Radio-Loud Quasars} \\ \hline
0137+012  & G & 0.258&  24/09/92 & 1.34 & 2.0 & 5$\times$1800& 3$''$&3662-7121 &5$''$E, 155$^{\circ}$	     	 \\
          & N & 0.258&  26/09/92 & 1.70 & 1.8 &  1$\times$900& 3$''$&3659-7121 &zero offset           	 \\
0736+017  & G & 0.191&  19/03/93 & 1.20 & 2.9 & 5$\times$1800& 3$''$&3530-7004 &5$''$S, 90$^{\circ}$	           \\
          & G & 0.191&  20/03/93 & 1.19 & 1.4 & 4$\times$1800& 3$''$&3530-7004 &5$''$S, 90$^{\circ}$	           \\
          & N & 0.191&  19/03/93 & 1.40 & 3.1 &  1$\times$900& 3$''$&3530-6989 &zero offset              \\     
          & N & 0.191&  20/03/93 & 1.17 & 1.4 & 1$\times$1200& 3$''$&3530-6989 &zero offset	        \\
1020$-$103& G & 0.197&  19/03/93 & 1.41 & 2.9 & 3$\times$1800& 3$''$&3530-7004 &3.5$''$W, 3.5$''$N, 45$^{\circ}$  	 \\
          & G & 0.197&  20/03/93 & 1.38 & 1.4 & 4$\times$1800& 3$''$&3530-7004 &3.5$''$W, 3.5$''$N, 45$^{\circ}$  	 \\
          & N & 0.197&  19/03/93 & 1.36 & 2.9 &  1$\times$900& 3$''$&3530-6989 &zero offset           	 \\
2201+315  & G & 0.298&  25/09/92 & 1.09 & 1.8 & 5$\times$1800& 3$''$&3665-7121 &5$''$N, 84$^{\circ}$	         \\
          & N & 0.298&  25/09/92 & 1.00 & 1.8 & 1$\times$1800& 3$''$&3665-7121 &zero offset  	     	 \\
2247+140  & G & 0.237&  24/09/92 & 1.26 & 1.2 & 5$\times$1800& 3$''$&3665-7124 &4$''$E, 3$''$S, 38.7$^{\circ}$    	 \\
          & N & 0.237&  24/09/92 & 1.06 & 1.2 & 1$\times$1800& 3$''$&3665-7124 &zero offset      \\ \hline
\multicolumn{10}{c}{\it Radio Galaxies} \\ \hline
1330+022  & G & 0.215&  20/03/93 & 1.18 & 1.4 & 5$\times$1800& 3$''$&3548-7004 &2.5$''$N, 2.5$''$E, 135$^{\circ}$ 	 \\
          & N & 0.215&  19/03/93 & 1.18 & 3.1 & 2$\times$1800& 3$''$&3548-7004 &zero offset           	 \\
2141+279  & G & 0.215&  26/09/92 & 1.03 & 1.8 & 4$\times$1800& 3$''$&3542-7124 &4$''$E, 0$^{\circ}$   	     	 \\
          & N & 0.215&  26/09/92 & 1.13 & 1.6 & 4$\times$1800& 3$''$&3542-7124 &zero offset           	 \\ \hline

\end{tabular}
\end{table*}


\begin{table*}
\small
\caption{Galaxies observed with ISIS on the 4.2~m William Herschel
Telescope on La Palma. Unlike the observations at Kitt Peak, nuclear
spectra were not obtained for all objects observed with the WHT. Where
such spectra are available, either from Kitt Peak or WHT observations,
they are listed here and displayed in Figure~2 along with the
off-nuclear (galaxy) spectrum. `M4M' denotes that the nuclear spectrum
was taken with the Mayall 4-m Telescope at Kitt Peak and `WHT' that
the spectrum was obtained using ISIS on the William Herschel
Telescope. Columns are as follows: (1) source name, listed in RA
order; (2) type of spectrum, `G' $=$ off-nuclear (galaxy) spectrum,
`N' $=$ nuclear spectrum; (3) redshift; (4) date of observation
(dd/mm/yy); (5) average airmass during integration period; (6) average
seeing during observations (arcsec); (7) number and length (in
seconds) of consecutive exposures; (8) slit width ($2''$ for all data
except the six nuclear spectra obtained at Kitt Peak); (9) wavelength
range determined by grating settings; (10) `join' wavelength at which
red and blue spectra have been spliced together; (11) positional
offset of slit centre from quasar/galaxy centroid and PA of slit
(measured East from North).}

\begin{tabular}{lcccccrcccl}

\hline
\footnotesize
Source  &Type&z&Date&Airmass&Seeing&Exposure&Slit &$\lambda$-range&$\lambda$ join&\multicolumn{1}{c}{Slit offset} \\
        &    & &    &       &($''$)&(sec)   &width&(\AA)         &(\AA)         &\multicolumn{1}{c}{\& PA}        \\ \hline
\multicolumn{11}{c}{\it Radio-Quiet Quasars} \\ \hline
0054+144 &  G   & 0.171 &   01/09/94& 1.30 &0.9 & 4$\times$1800& 2$''$& 3240-8700& 6050 &3.5$''$N, 4$''$W, 228$^{\circ}$   \\
         &N(M4M)& 0.171 &   26/09/92& 1.60 &1.6 &  1$\times$900& 3$''$& 3659-7121&   -- &zero offset                       \\
0157+001 &  G   & 0.164 &   16/11/93& 1.15 &1.6 & 4$\times$1800& 2$''$& 3456-9060& 6100 &5$''$N, 107.5$^{\circ}$           \\
         &N(M4M)& 0.164 &   25/09/92& 1.55 &1.9 & 1$\times$1200& 3$''$& 3662-7121&   -- &zero offset                       \\
0204+292 &  G   & 0.109 &   17/11/93& 1.03 &1.0 & 5$\times$1800& 2$''$& 3456-9060& 6100 &4.5$''$W, 0$^{\circ}$             \\
0244+194 &  G   & 0.176 &   02/09/94& 1.06 &1.7 & 4$\times$1800& 2$''$& 3240-8700& 6050 &5$''$W, 180$^{\circ}$             \\
1549+203 &  G   & 0.250 &   30/03/95& 1.05 &1.2 & 4$\times$1800& 2$''$& 3450-8991& 6100 &4$''$W, 3$''$N, 228$^{\circ}$     \\
1635+119 &  G   & 0.146 &   13/05/94& 1.12 &1.3 & 5$\times$1800& 2$''$& 3240-8700& 6050 &4.75$''$W, 1$''$S, 172$^{\circ}$  \\
2215$-$037& G   & 0.241 &   01/09/94& 1.60 &0.9 & 3$\times$1800& 2$''$& 3240-8700& 6050 &4.25$''$W, 3$''$N, 210$^{\circ}$  \\
2344+184 &  G   & 0.138 &   02/09/94& 1.20 &1.6 & 4$\times$1800& 2$''$& 3240-8700& 6050 &3$''$N, 4$''$W, 215$^{\circ}$     \\
         &N(M4M)& 0.138 &   24/09/92& 1.15 &2.5 & 1$\times$1800& 3$''$& 3662-7121&   -- &zero offset                       \\ \hline
\multicolumn{11}{c}{\it Radio-Loud Quasars} \\ \hline
0736+017 &  G   & 0.191 &   16/11/93& 1.14 &1.0 & 5$\times$1800& 2$''$& 3456-9060& 6100 &3.5$''$S, 3.5$''$W, 135$^{\circ}$ \\
         &N(M4M)& 0.191 &   19/03/93& 1.40 &3.1 &  1$\times$900& 3$''$& 3530-6989&   -- &zero offset                       \\
1004+130 &  G   & 0.240 &   30/03/95& 1.20 &1.5 & 4$\times$1800& 2$''$& 3450-8991& 6100 &3.5$''$N, 3.5$''$W, 43$^{\circ}$  \\
1217+023 &  G   & 0.240 &   11/05/94& 1.20 &0.8 & 4$\times$1800& 2$''$& 3240-8700& 6050 &4$''$E, 3$''$N, 334$^{\circ}$     \\
2135$-$147& G   & 0.200 &   02/09/94& 1.60 &1.2 & 4$\times$1800& 2$''$& 3240-8700& 6050 &3.5$''$W, 3.5$''$S, 315$^{\circ}$ \\
2141+175 &  G   & 0.213 &   13/05/94& 1.60 &1.0 & 3$\times$1800& 2$''$& 3240-8700& 6050 &5$''$N, 290$^{\circ}$             \\
         &N(WHT)& 0.213 &   13/05/94& 1.30 &1.1 &  1$\times$900& 2$''$& 3240-8700& 6050 &zero offset                       \\
2247+140 &  G   & 0.237 &   16/11/93& 1.04 &1.6 & 2$\times$1800& 2$''$& 3456-9060& 6100 &4$''$E, 3$''$S, 38.7$^{\circ}$    \\
         &N(M4M)& 0.237 &   24/09/92& 1.06 &1.2 & 1$\times$1800& 3$''$& 3665-7124&   -- &zero offset                       \\
2349$-$014& G   & 0.173 &   04/09/94& 1.35 &0.7 & 4$\times$1800& 2$''$& 3240-8700& 6050 &1$''$S, 5$''$W, 324$^{\circ}$     \\ \hline
\multicolumn{11}{c}{\it Radio Galaxies} \\ \hline
0230$-$027& G   & 0.239 &   02/09/94& 1.30 &2.0 & 2$\times$1800& 2$''$& 3240-8700& 6050 &3.5$''$S, 3.5$''$W, 320$^{\circ}$ \\
         &N(WHT)& 0.239 &   01/09/94& 1.20 &1.5 & 1$\times$1800& 2$''$& 3240-8700& 6050 &zero offset                       \\
0345+337 &  G   & 0.244 &   02/09/94& 1.30 &3.0 & 4$\times$1800& 2$''$& 3240-8700& 6050 &5$''$N, 0.5$''$W, 265$^{\circ}$   \\
0917+459 &  G   & 0.174 &   19/11/93& 1.10 &1.7 & 5$\times$1800& 2$''$& 3456-9060& 6100 &5$''$N, 90$^{\circ}$   	   \\
1215$-$033& G   & 0.184 &   13/05/94& 1.19 &0.9 & 3$\times$1800& 2$''$& 3240-8700& 6050 &4$''$S, 3$''$E, 50$^{\circ}$      \\
1334+008 &  G   & 0.299 &   30/03/95& 1.60 &2.0 & 2$\times$1800& 2$''$& 3450-8991& 6100 &3.5$''$W, 3.5$''$S, 135$^{\circ}$ \\
2141+279 &  G   & 0.215 &   02/09/94& 1.20 &1.0 & 4$\times$1800& 2$''$& 3240-8700& 6050 &4.5$''$N, 2$''$E, 296$^{\circ}$   \\
         &N(M4M)& 0.215 &   26/09/92& 1.13 &1.6 & 4$\times$1800& 3$''$& 3542-7124&   -- &zero offset                       \\ \hline

\end{tabular}
\end{table*}

\section{Observations} 

In order to asssess the feasibility of our observing strategy initial
observations of 11 of the nearest and brightest objects in the sample
were carried out using the Mayall 4-m Telescope at Kitt Peak National
Observatory.  This was followed by a larger programme of observations
using the 4.2-m William Herschel Telescope (WHT), part of the Isaac
Newton Group of telescopes on La Palma.

Tables 2 and 3 list the observations made on the Mayall 4-m Telescope
and the WHT respectively. Six objects (0054$+$144, 0157$+$001,
0736$+$017, 2141$+$279, 2247$+$140 and 2344$+$184) were observed with
both telescopes in order to provide a check for consistency between
the two sets of observations. The spectra for these objects are
discussed in more detail in Section~6.

\subsection{The Mayall 4-m Telescope}

Observations of 11 objects were carried out with the
R. C. Spectrograph on the Mayall 4-m Telescope at Kitt Peak in 1992
September and 1993 March (Table~1). The long-slit spectrograph uses a
Tektronix $2048\times2048$ chip with 24-$\mu$m pixels, designated
T2KB.  The slit width was set to 3$''$ and the instrument was
configured to give a spectral resolution of 1.9\AA/pixel and a spatial
resolution of 0.69$''$/pixel.

The slit was first centred on the quasar (or, if the object was a
radio galaxy, on the galaxy centroid), before being rotated and offset
to the desired position, usually 5~arcseconds off-nucleus.  

To allow the removal of cosmic rays five 1800-second off-nuclear
exposures were obtained for each object, along with a shorter exposure
of the quasar itself.  The spectra typically spanned wavelengths from
3500\AA~ to 6000\AA, although the precise range varied from object to
object according to redshift; details for individual objects are given
in Table~2.  Data reduction was carried out using standard {\sc iraf}
routines.

\subsection{Observations with ISIS on the WHT}

After the two runs at Kitt Peak had demonstrated that galaxy light
could indeed be separated from that of the quasar and that useful
spectra could be obtained, further observations were made of 22
objects (6 of which had already been observed at Kitt Peak) with the
Intermediate Dispersion Spectroscopic and Imaging System (ISIS) at the
4.2~m William Herschel Telescope (WHT) on La Palma. ISIS uses a
dichroic mirror to split the incoming light into red and blue beams
which are then treated separately in spectrographs which have been
optimised for the appropriate wavelength ranges
($3000\rightarrow6000$\AA~ for the blue arm and
$5000\rightarrow10000$\AA~ for the red). This arrangement allows a
larger wavelength coverage than is possible with the single
spectrograph on the Mayall 4-m, enabling us to extend our spectra
further into the red and thus giving greater scope for constraining
models of spectrophotometric evolution.

The data were obtained in four separate observing runs in 1993
November, 1994 May, 1994 September and 1995 March. The instrumental
set-up differed slightly between the runs, resulting in variations in
the wavength ranges obtained. For the two sessions in 1994 the detectors
in both the blue and red arms were Tektronix (`Tek') CCDs, with 24-$\mu$m
pixels. In November 1993 and March 1996 the Tektronix chip in the red arm
was replaced by an EEV~P88300 chip, with 22.5-$\mu$m pixels, which
allowed a slightly larger wavelength coverage.  R158 gratings were
used in each arm, giving a spectral resolution of 2.88\AA~pixel$^{-1}$
for the blue Tek chip and 2.90 (Tek) or 2.72\AA~pixel$^{-1}$ (EEV) for
the red chips.

As with the observations at Kitt Peak the slit was first centred on
the quasar nucleus, or the optical peak of the radio galaxy, and then
offset to the desired position, 5$''$ from the quasar, and rotated to
be at right angles to the direction of the offset. The slit width was
set to 2$''$, placing the inner edge of the slit at least 4$''$ from
the quasar position. 

Once again exposures were limited to 1800 seconds duration. Whenever
possible we aimed to obtain five such frames per object, giving a
total on-source exposure time of 2.5 hours. On-nuclear quasar spectra
were also taken when time permitted.  Data reduction was carried out
using the {\sc figaro} package, part of the Starlink suite of
astronomical software.

\subsubsection{Calibration and splicing of red and blue spectra}

The galaxies observed with the WHT tend to be fainter and/or at
greater redshifts than those observed on the Mayall 4-m at Kitt Peak,
and so additional care needed to be taken during the reduction of the
WHT spectra. In particular, the procedure employed to optimize the
extraction of the off-nuclear spectrum from the CCD frame means that
the final flux calibration is only relative and not absolute -
comparisons with the absolute fluxes obtained on the Mayall 4-m are
therefore not meaningful.

The use of the red and blue arms of ISIS, whilst extending the
wavelength coverage significantly, also introduces its own special
problems.  The large wavelength range made available by ISIS means
that the spectra are affected by several prominent atmospheric
emission features (Figure~1) which must be removed. Of these the
strongest are the two oxygen lines at 5577 and 6300\AA, the sodium D
line at 5890\AA, and the series of OH bands at wavelengths
$>6500$\AA. Sky lines were removed by fitting a third order polynomial
to two empty strips of sky on either side of the target spectrum but
the removal process sometimes left a residual imprint of these
features and this constitutes a major source of noise in some of the
fainter spectra in our sample. Where this is the case the affected
regions are mentioned in the description of the individual spectrum.

\begin{figure}
\setcounter{figure}{0}
\vspace{4.5cm}
\includegraphics{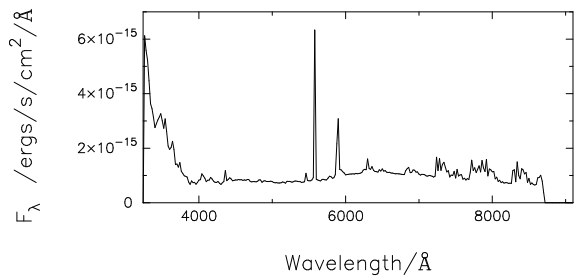}

\caption{Sky spectrum obtained at the WHT in September 1994. For the
fainter targets in the sample the process of sky subtraction often
leads to residuals around the positions of strong sky emission
features such as the two lines of neutral oxygen at 5577 and 6300\AA,
the sodium D line at 5890\AA, and the series of OH bands at wavelengths
$>6500$\AA. }
\end{figure}

A second problem involves the splicing together of the two spectra
from the red and blue arms of the instrument to give continuous
wavelength coverage from 3200\AA~ to 9000\AA. 

The reflection and transmission responses of the ISIS dichroic cross
at $\sim6100$\AA~ and the instrument settings were designed to ensure
a large overlap between the wavelength coverages of the two arms so
that the full reflection/transmission curves could be followed in the
blue and the red spectra respectively.  However, the faintness of many
of the spectra effectively causes the flux calibration to fail at the
red end of the blue spectrum and the blue end of the red, producing an
artificial `lump' in the spliced spectrum at the crossover point
(6100\AA~ or 6050\AA, depending on the exact instrumental settings at
the time of the observations - see column~8 in Table~3).

Accordingly, in plotting the WHT spectra in Figure~2 we have blanked out
the data points for 100\AA~ on either side of the join region and
replaced them with an averaged bridging section. We have masked out
this section of the data in all subsequent modelling due to its
inherrent unreliability (see Nolan et al. 2000). 

Due to the luminosity of the quasars themselves the on-nuclear spectra
obtained with the WHT do not suffer from this problem and the
red and blue sections have been spliced together directly with no
noticeable discontinuity in flux at the join.

\section{Results}

The galaxy spectra themselves are displayed in Figure~2 along with
nuclear spectra for the same object, where such are available. All the
data have been smoothed to 10-\AA~ bins, and the WHT data are
displayed with the region linking the individual red and blue spectra
blanked out as described above. The expected position of various
stellar absorption features, including the 4000\AA~ break are
indicated by dotted lines in the off-nuclear spectra and the
wavelengths of redshifted [O{\sc iii}]$\lambda5007$ and H$\alpha$ are
also marked.

\subsection{Data quality}

A cursory examination of Figure~2 is enough to show that the quality
of the off-nuclear spectra varies considerably from object to
object. In the two initial observing runs at Kitt Peak priority was
given to nearby objects with bright, prominent host galaxies, and the
eleven spectra obtained with the Mayall 4-m enjoy a high
signal-to-noise ratio. By contrast, the objects observed in later runs
on the WHT tend to have larger redshifts and the signal-to-noise in
many of these spectra is correspondingly lower due to the rapid
reduction in surface brightness and increase in galactocentric radius
with $z$.

\subsection{Degree of quasar contamination}

The primary goal of the observing program - to obtain spectra from the
stellar component of the host whilst avoiding scattered emission from
the active nucleus - appears to have been satisfied to a large
extent. This can be demonstrated most easily by comparing the
off-nuclear (galaxy) and nuclear spectra for the same object in Figure
2. The quasar spectra show prominent broad lines, notably those of
H$\alpha~\lambda 6563$ and H$\beta~\lambda 4861$ , with an increase in
flux towards shorter wavelengths (particularly when the `blue bump'
continuum feature begins to emerge at $\lambda_{rest} \leq 5000$\AA).

Whilst emission lines do occur in the off-nuclear spectra, they tend
to be relatively narrow forbidden lines such as [O{\sc
iii}]$\lambda \lambda \lambda 4363,4959,5007$. Where permitted lines
occur, they lack the extremely broad profiles seen in the quasar
nuclei, and are thus unlikely to result from scattering-induced
contamination by nuclear light.  In many cases the line ratios in the
off-nuclear spectra also differ from those measured in the quasars,
indicating that the emission arises under different conditions than
those prevailing in the active nucleus.

However, in spectra obtained under conditions of poor seeing, there is
likely to be some degree of nuclear contamination and this will be
exacerbated if the object was observed far from the zenith, where
differential atmospheric refraction may also lead to signifcantly more
contamination at the blue end of our wavelength range than in the
red. Although the extent of such contamination is difficult to measure
directly, Tables 2 and 3 list the atmospheric seeing and airmass at
the time each spectrum was obtained, to allow a rough assessment of
the problem to be made.

\subsection{Spectral features}

\begin{table}
\caption{Strength of the 4000\AA~break feature in the host galaxy
spectra (defined as the ratio of the flux density between 4050 and
4250\AA~ to that between 3750 and 3950\AA~ in the object's rest frame,
and measured in terms of $F_{\nu}$). M4M denotes the Mayall 4-m Telescope
at Kitt Peak; WHT denotes the 4.2-m William Herschel Telescope on La
Palma. 3-sigma errors were estimated from the scatter in the flux
density across the two reference regions.}

\begin{tabular}{ccc}

\hline
\footnotesize
IAU  & Telescope & Break  \\
name &           & strength \\ \hline
\multicolumn{3}{c}{\it Radio-Quiet Quasars} \\ \hline
0007$+$106  &  M4M & $ 1.4  \pm  0.1$ \\                      
0054$+$144  &  M4M & $ 1.3  \pm  0.1$ \\                      
            &  WHT & $ 1.3  \pm  0.1$ \\  
0157$+$001  &  M4M & $ 1.3  \pm  0.1$ \\         
            &  WHT & $ 1.3  \pm  0.1$ \\  
0204$+$292  &  WHT & $ 1.6  \pm  0.1$ \\                      
0244$+$194  &  WHT & $ 0.8  \pm  0.2$ \\  
1549$+$203  &  WHT & $ 2.0  \pm  0.1$ \\              
1635$+$119  &  WHT & $ 1.7  \pm  0.1$ \\              
2215$-$037  &  WHT & $ 1.4  \pm  0.1$ \\  
2344$+$184  &  M4M & $ 1.7  \pm  0.1$ \\  
            &  WHT & $ 1.4  \pm  0.1$ \\ \hline                  
\multicolumn{3}{c}{\it Radio-Loud Quasars} \\ \hline
0137$+$012  &  M4M & $ 1.7  \pm  0.1$ \\                   
0736$+$017  &  M4M & $ 1.4  \pm  0.1$ \\              
            &  WHT & $ 1.8  \pm  0.1$ \\
1004$+$130  &  WHT & $ 1.1  \pm  0.1$ \\    
1020$-$103  &  M4M & $ 1.7  \pm  0.1$ \\              
1217$+$023  &  WHT & $ 1.1  \pm  0.1$ \\              
2135$-$147  &  WHT & $ 2.0  \pm  0.3$ \\             
2141$+$175  &  WHT & $ 0.9  \pm  0.2$ \\
2201$+$315  &  M4M & $ 1.4  \pm  0.1$ \\             
2247$+$140  &  M4M & $ 1.9  \pm  0.1$ \\             
            &  WHT & $ 2.0  \pm  0.1$ \\
2349$-$014  &  WHT & $ 1.5  \pm  0.1$ \\ \hline              
\multicolumn{3}{c}{\it Radio Galaxies} \\ \hline
0230$-$027  &  WHT & $ 1.9  \pm  0.4$ \\                  
0345$+$337  &  WHT & $ 1.7  \pm  0.4$ \\                         
0917$+$459  &  WHT & $ 2.0  \pm  0.1$ \\            
1215$-$033  &  WHT & $ 1.9  \pm  0.2$ \\
1330$+$022  &  M4M & $ 1.6  \pm  0.1$ \\            
1334$+$008  &  WHT & $ 1.1  \pm  0.2$ \\                   
2141$+$279  &  M4M & $ 2.2  \pm  0.1$ \\
            &  WHT & $ 1.3  \pm  0.2$ \\ \hline                        

\end{tabular}
\end{table}

The characteristic shape of the stellar continuum, including features
such as the 4000\AA~ break and various stellar absorption lines, is
easily recognisable in most cases.  The 4000\AA~ break is particularly
important for comparison between data and models of the stellar
population. Since it covers a large wavelength interval it is
relatively insensitive to the effects of instrument resolution and
noise (Hamilton 1985). The break amplitude (defined as the ratio of
the average flux density between rest-frame 4050 and 4250\AA~ to that
between 3750 and 3950\AA) is therefore widely used as a tracer of
spectral evolution ({\it eg} Bruzual 1983). The discontinuity results
from the combined effect, shortwards of 4000\AA, of lines of several
elements heavier than helium, in a variety of ionization states, along
with the crowding of higher order Balmer lines. If a significant
population of massive young stars is present the enhanced degree of
ionization causes the feature to weaken; it is most prominent in the
spectrum of a well-established stellar population in which the most
massive stars have had time to evolve away from the main sequence.
Hence the 4000\AA~break is sensitive to both spectral type and
metallicity, although if we assume a constant metallicity (a
reasonable assumption for a particular galaxy at a fixed radius) it
becomes a good indicator of the mean age of the local stellar
population. The strength of this feature, as measured in each of the
current spectra, is listed in Table~4. Other stellar absorption
features, such as G band (4300 to 4320\AA) and the Mg Ib (5173\AA) and
Fe$\lambda5270$ lines, are also clearly present in many of the
spectra.

Several of the off-nuclear spectra ({\it eg} 0054+144, 2201+315),
despite showing a clear break at 4000\AA~ and a continuum longwards of
this wavelength which can be fitted extremely well by a passively
ageing stellar population (Nolan et al. 2000), also display a
contribution from a component which rises steeply towards the
blue. The slope of this feature closely resembles that of a quasar
SED, and its presence is often (but not always) accompanied by
emission lines characteristic of quasar nuclei, suggesting that it is
in fact scattered light from the quasar itself, the result either of
atmospheric scattering or (since the feature is not always
correlated with poor observing conditions) of scattering within the
ISM of the host galaxy. Another possibility is that the blue excess
indicates the presence of a substantial population of young stars
within the host galaxy. If this latter case were true then it would
pose a serious problem for the unification of RLQs and RGs since none
of the radio galaxies display such a component. The issue of excess
blue continuum is raised on a case-by-case basis in the following
section, but a full discussion in the light of detailed stellar
population synthesis modelling is deferred to the companion paper by
Nolan et al. (2000).

The average values of the 4000\AA~break strength for the three types
of object in the sample are 1.4 (RQQ hosts), 1.5 (RLQ hosts) and 1.7
(RGs). These are all somewhat lower than the value of $\sim
2.0$ measured for local inactive elliptical galaxies by Hamilton
(1985), but we note that there is a wide scatter in our sample and
that the lowest values are all associated either with spectra in which
the signal to noise is particularly poor ({\it eg} 0244+292, 1004+130,
1334+008) or those which clearly show an extra source of continuum
emission at short wavelengths ({\it eg} the quasars 0054+144,
1217+023, 2141+175). The cleanest spectra generally have break
strengths which are consistent with those seen in `normal'
well-established galaxies.

As a final caveat we note that these spectra tell us only about the
stellar composition of the region of the galaxy covered by the slit -
we cannot, for example, rule out the presence of a significantly
different stellar population closer to the nucleus, or concentrated in
clumps which the slit happens to avoid.  However, the slit has a width
of at least 2~arcsec and, as can be seen from the contour plots in
Figure 2, its length cuts across a significant fraction of the galaxy
in the transverse direction. The area covered often amounts to several
square arcseconds (equivalent to several tens of square kiloparsecs at
typical redshifts) and therefore represents a good general sample of
the outer regions of the host.

\section{Individual objects}

Objects are listed under their IAU names (alternative names are listed
in Table~1), in order of increasing right ascension.  The
classification of each object as either an RQQ, an RLQ or a radio
galaxy is indicated in parentheses, along with the name(s) of the
telescope(s) on which off-nuclear spectra were obtained. The form of
the spectrum is described, noting any peculiar features as well as the
presence or otherwise of a 4000\AA~ break at the expected observed
wavelength ($\lambda_{obs}$). We also note the morphology of the
galaxy (disc or elliptical) based on its surface brightness profile in
the $K$ or $R$-band continuum images by Taylor et al. (1996)
($K$-band; UKIRT) or McLure et al. (1999) and Dunlop et al. (2000)
($R$-band; HST).  For a more detailed description of previous imaging
studies of the host galaxies see Dunlop et al. (1993) (RLQs and RQQs)
or Taylor et al. (1996) (RGs).

\noindent
{\bf 0007$+$106 (RQQ; M4M):} the nuclear spectrum of this radio-quiet
quasar shows prominent broad H$\alpha \lambda 6583$ and H$\beta
\lambda 4861$ emission as well as narrower forbidden line emission
from species including [0{\sc iii}] and [Fe{\sc vii}]. The off-nuclear
spectrum of the host galaxy has a high signal-to-noise and displays
little sign of contamination from the quasar: the contribution from
emission lines is very small, and the 4000\AA~ break is clearly
visible (redshifted to 4356\AA) despite the rapid increase in quasar
flux towards the blue end of the spectrum, which would tend to mask
the break if scattering were significant. G band and Mg {\sc i}b
absorption are also present. We note that the slit crosses the optical
arc-like structure to the north of the quasar which Hutchings et
al. (1984) suggest may be a spiral arm. Previous spectroscopy of this
region by Green, Williams \& Morton (1978) showed narrow emission
lines (with different line ratios from those in the nucleus) and a red
continuum which they attribute to starlight. However, this region only
constitutes a small fraction of the galaxy light intercepted by the
slit in the current observations. Taylor et al. (1996) find that an
exponential disc profile provides a good fit to the NIR surface
brightness distribution of the galaxy. (Morphology: disc; Taylor et
al. 1996.)

\noindent
{\bf 0054$+$144 (RQQ; M4M \& WHT):} quasar continuum emission
dominates the nuclear spectrum of this RQQ, although H$\beta \lambda
4861$ and [O{\sc iii}]$\lambda\lambda\lambda 4363,4959,5007$ lines are
visible. These lines are not prominent in either the Mayall 4-m or WHT
off-nuclear spectra of the host (particularly the latter spectrum) but
bluewards of the (relatively weak) 4000\AA~ break (at $\lambda_{obs} =
4684$\AA) the galaxy spectrum displays a marked increase in flux, very
similar in form to that displayed by the quasar itself. This component
is seen in both off-nuclear spectra, which were taken at different
times, under different seeing conditions, and used different slit
positions, so it is not clear whether we are seeing quasar light which
is being scattered into our line of sight either by the atmosphere or
by the interstellar medium of the host galaxy, or whether the blue
excess is due to a population of young stars. G band, Mg {\sc i}b and
Fe5270 absorption features appear in the Mayall 4-m spectrum.  McLure
et al. (1999) classify the galaxy as an elliptical: its light profile
is very well described by an $r^{1/4}$ law and the galaxy itself is
quite red ($R-K=3.14$), but a tidal interaction with a nearby
companion is suggested by the extension to the NW of the nucleus
(Dunlop et al. 1993). (Morphology: elliptical; McLure et al. 1999.)

\noindent
{\bf 0137$+$012 (RLQ; M4M):} the 4000\AA~ break in this object is quite
clear (at $\lambda_{obs} = 5032$\AA) and there is little evidence of
nuclear emission lines. (Morphology: elliptical; McLure et al. 1999.)

\noindent
{\bf 0157$+$001 (RQQ; M4M \& WHT):} the off-nuclear spectrum obtained
with the Mayall 4-m does not appear to be strongly contaminated by
emission from the nucleus and the 4000\AA~ break is visible at
$\lambda_{obs} = 4656$\AA. The G band absorption feature is also
present. However, in the spectrum taken with the WHT using a similar
slit position, there appears to be a significant contribution from the
quasar continuum bluewards of the break. The signal:noise ratio in the
WHT spectrum is much reduced longwards of 7500\AA~ due to residuals
from the subtracted OH bands. In both cases the slit intercepts the
prominent tidal arm which extends north and NW of the nucleus
(MacKenty \& Stockton 1984) and is known to contain several emission
line regions (Stockton \& MacKenty 1987), although emission lines are
not strongly evident in the integrated spectra presented
here. Previous spectroscopy of this structure by Heckman et al. (1984)
showed a velocity difference of $300\pm200$~km~s$^{-1}$ between the
arm and the quasar itself. McLure et al. (1999) find that the
underlying smooth $R$-band continuum light is well described by an
$r^{1/4}$-law, though conceivably this might be another result of the
tidal interaction.  (Morphology: elliptical; McLure et al. 1999.)

\noindent
{\bf 0204$+$292 (RQQ; WHT):} the 4000\AA~break ($\lambda_{obs}
=4436$\AA) is quite strong in this WHT spectrum, but residuals from
oxygen and OH features from the sky spectrum are also present.
(Morphology: disc; Taylor et al. 1996)

\noindent
{\bf 0230$-$027 (RG; WHT):} the off-nuclear spectrum of this small, faint
radio galaxy suffers from low signal to noise and residual sky
features are visible. However a break can be seen at $\lambda_{obs} =
4952$\AA. The nuclear spectrum of this object is also dominated by
starlight, although narrow emission lines are present.
(Morphology: elliptical; Dunlop et al. 2000.)

\noindent
{\bf 0244$+$194 (RQQ; WHT):} the signal to noise in this off-nuclear
spectrum is very low. Apart from residual sky features there is some
evidence for a drop in the continuum level around $\lambda_{obs} =
4704$\AA~ (the expected position of the 4000\AA~break), but also for a
blue component shortwards of this, perhaps indicative of nuclear
contamination. (Morphology: elliptical; McLure et al. 1999.)

\noindent
{\bf 0345$+$337 (RG; WHT):} the spectrum is noisy, but a break feature
appears to be present at $\lambda_{obs} =4936$\AA. The WHT slit
intercepts a bright knot NW of the quasar which Taylor et al. (1996)
suggest may be an embedded companion galaxy. (Morphology: elliptical;
McLure et al. 1999.)

\noindent
{\bf 0736$+$017 (RLQ; M4M \& WHT):} the WHT spectrum, though noisier
and suffering from OH-band residuals at long wavelengths, agrees well
with the spectrum obtained previously at Kitt Peak. The 4000\AA~break
is quite clear at $\lambda_{obs} =4764$\AA~ and weak [O{\sc iii}],
H$\alpha$ and H$\beta$ features can also be seen, though they appear
to be narrower than those in the nuclear spectrum. The galaxy itself
is highly disturbed, but McLure et al. (1999) fit an $r^{1/4}$-law
profile to the smooth component of the $R$-band continuum. (Morphology: 
elliptical; McLure et al. 1999.)

\noindent
{\bf 0917$+$459 (RG; WHT):} the off-nuclear spectrum is pure stellar
continuum, with a strong break at $\lambda_{obs} =4696$\AA. The galaxy
isophotes are complex but McLure et al. (1999) find the underlying
distribution to be well described by an $r^{1/4}$-law.  (Morphology:
elliptical; McLure et al. 1999.)

\noindent
{\bf 1004$+$130 (RLQ; WHT):} OH-band residuals are the only prominent
feature of this off-nuclear spectrum, with little evidence for either
emission lines or a 4000\AA~break (at $\lambda_{obs} =4960$\AA). The
absence of a strong break may reflect the fact that the elliptical
host galaxy of this quasar is known to possess unusual `spiral'
features close to the nucleus (McLure et al. 1999), perhaps
indicating the presence of a significant population of young
stars. Stockton \& MacKenty (1987) note that there is no significant
extended [O{\sc iii}] emission in this object. (Morphology:
elliptical; McLure et al. 1999.)

\noindent
{\bf 1020$-$103 (RLQ; M4M):} despite the presence of stellar
absorption features such as G band and a 4000\AA~break at
$\lambda_{obs} =4788$\AA, the off-nuclear spectrum also contains many
emission lines as well as a blue excess, which may indicate a
significant contribution from scattered quasar light (the seeing was
quite poor for much of the observations). Dunlop et al. (1993) note
that the quasar is off-centre and that the galaxy isophotes appear to
be swept back towards the SW, providing evidence for disturbance in
this object. (Morphology: elliptical; Dunlop et al. 1999.)

\noindent
{\bf 1215$-$033 (RG; WHT):} this spectrum displays only a weak break at
$\lambda_{obs} =4736$\AA~. OH-band residuals dominate longwards of
7500\AA. (Morphology: elliptical; Dunlop et al. 2000.)

\noindent
{\bf 1217$+$023 (RLQ; WHT):} a weak break feature is present at
$\lambda_{obs} =4960$\AA~ but shortwards of this the spectrum is
dominated by a component which rises towards the blue possibly
indicating scattered nuclear continuum (there is however little
evidence for accompaning nuclear line emission, and the seeing during
the observations was excellent). At the red end, poor subtraction of OH
bands has reduced the signal to noise ratio of the spectrum.
(Morphology: elliptical; Dunlop et al. 2000.)

\noindent
{\bf 1330$+$022 (RG; M4M):} weak [O{\sc iii}] lines occur in the
off-nuclear spectrum of this radio galaxy and the stellar continuum
shows G band absorption and a strong 4000\AA~break feature at
$\lambda_{obs} =4860$\AA. By contrast, the nuclear spectrum shows
evidence for broad H$\beta \lambda 4861$ and the less prominent break
may indicate a contribution from a quasar-type continuum or perhaps a
nuclear starburst region.  (Morphology: elliptical; Dunlop et
al. 2000.)

\noindent
{\bf 1334$+$008 (RG; WHT):} a very noisy spectrum obtained under poor
conditions, the underlying continuum is confused by many residual sky
features. The apparent increase in flux shortwards of 4500\AA~ almost
certainly reflects a failure of the flux calibration at very low light
levels. However, there is some evidence for a break at the expected
wavelength of $\lambda_{obs} =5196$\AA. The slit intercepts one of the
secondary nuclei reported by Taylor et al. (1996). (Morphology:
elliptical; Taylor et al. 1996.)

\noindent
{\bf 1549$+$203 (RQQ; WHT):} the extended nebulosity around this RQQ,
though confused by a foreground galaxy cluster, shows a strong break
feature at the expected wavelength of $\lambda_{obs} =5000$\AA~ and
little evidence for emission lines. OH-band residuals add to the noise
levels at long wavelengths. (Morphology: elliptical; Dunlop et al. 2000.)

\noindent
{\bf 1635$+$119 (RQQ; WHT):} the 4000\AA~break is clearly visible at
$\lambda_{obs} =4584$\AA, but residual sky features degrade the
quality of the spectrum towards the red end. (Morphology: elliptical;
McLure et al. 1999.)

\noindent
{\bf 2135$-$147 (RLQ; WHT):} generally low signal to noise with
prominent residuals due to sky features. However, there is weak
evidence for a break at $\lambda_{obs} =4800$\AA. The galaxy appears
to be disturbed and possesses a secondary nucleus to the SE of the
quasar (Stockton 1982) which may also be active (Hickson \& Hutchings
1987). The WHT slit intercepts a region to the SW of the quasar at
which Stockton \& MacKenty (1987) report extended [O{\sc iii}]
emission. Narrow H$\alpha$ is present in the current spectrum but the
[O{\sc iii}] lines fall within the join region where the signal is
unreliable. (Morphology: elliptical; Dunlop et al. 2000.)

\noindent
{\bf 2141$+$175 (RLQ; WHT):} low signal to noise and the presence of a
strong blue component may serve to mask any evidence of a
4000\AA~break in this object (expected at $\lambda_{obs}
=4852$\AA). The idea that this blue component originates as scattered
quasar light is bolstered by the possible presence of the H$\alpha$
line, but the issue is confused by strong sky residuals (although the
seeing was good, the airmass during the observations was relatively
high, so differential refraction might explain the presence of quasar
contamination at shorter wavelengths without requiring the presence of
a strong H$\alpha$ line). The elongated appearance of the galaxy is
apparently the result of an edge-on tidal arm consisting of old stars
(Stockton \& Farnham 1991). However, the WHT slit crosses the galaxy
to the NE of the quasar, where the starlight appears to follow a
bulge-dominated $r^{1/4}$-law (McLure et al. 1999a). (Morphology:
elliptical; McLure et al. 1999.)

\noindent
{\bf 2141$+$279 (RG; M4M \& WHT):} the off-nuclear spectrum taken at
Kitt Peak shows a weak [O{\sc iii}]$\lambda 5007$ line, a break
feature at $\lambda_{obs} =4860$\AA and Mg {\sc i}b absorption. The
WHT spectrum is much noisier, but generally consistent with the
features in the earlier data.  The nuclear spectrum of this radio
galaxy is also dominated by starlight, although prominent narrow lines
are present. (Morphology: elliptical; McLure et al. 1999.)

\noindent
{\bf 2201$+$315 (RLQ; M4M):} this object shows prominent G band
absorption but only a weak 4000\AA~break at $\lambda_{obs}
=5192$\AA, along with low-equivalent-width [O{\sc iii}] lines. A blue
component shortwards of the break is consistent with scattering of the
nuclear quasar continuum. (Morphology: ambiguous; Taylor et al. 1996.)

\noindent
{\bf 2215$-$037 (RQQ; WHT):} the spectrum is relatively noisy with
prominent residuals from all the bright sky features. The apparent
`hump' at $\sim 6000$\AA~ is an artifact that results from the sodium
D sky line coinciding with the beginning of the masking region that
links the red and blue halves of the the spectra from ISIS. There is
little evidence for a prominent break at the expected wavelength of
$\lambda_{obs} =4964$\AA. Both Hutchings et al. (1989) and McLure et
al. (1999b) classify the galaxy as a non-interacting elliptical
system. (Morphology: elliptical; Dunlop et al. 2000.)

\noindent
{\bf 2247$+$140 (RLQ; M4M \& WHT):} the galaxy's 4000\AA~break is
visible at $\lambda_{obs} =4948$\AA~ in the Mayall 4-m spectrum, along
with G band absorption. The data obtained on the WHT are consistent
with this, although residual sky features from oxygen and sodium D
lines and the OH bands are also present. (Morphology: elliptical;
McLure et al. 1999.)

\noindent
{\bf 2344$+$184 (RQQ; M4M \& WHT):} both the Mayall 4-m and WHT
spectra show a break at $\lambda_{obs} =4552$\AA~ as well as G band
absorption and are generally consistent with one another despite their
differing slit positions. The nuclear spectrum of this object is also
dominated by starlight; 2344$+$184 is one of the least powerful
quasars in the current study and technically qualifies as a Type 1
Seyfert galaxy. The surface brightness profile is that of a disc
galaxy, although a bulge dominates in the central regions (McLure et
al. 1999) and there is also evidence for a bar (Hutchings, Janson \&
Neff 1989).  (Morphology: disc; McLure et al. 1999.)

\noindent
{\bf 2349$-$014 (RLQ; WHT):} there is a weak break feature at
$\lambda_{obs} =4692$\AA~ and also evidence for weak H$\beta$, [O{\sc
iii}]$\lambda 5007$ and H$\alpha$ lines. The red end of the spectrum
is dominated by residual atmospheric OH band emission. (Morphology:
elliptical; McLure et al. 1999.)

\section{Comparison of Mayall 4-m and WHT results}

Six objects in the current sample (the RQQs 0054$+$144, 0157$+$001 and
2344$+$184, the RLQs 0736$+$017 and 2247$+$140, and the radio galaxy
2141$+$279) were observed with both the Mayall 4-m Telescope and the
WHT. This duplication allows us to check for systematic differences
between the spectra obtained with each instrument and to this end the
are replotted one above the other in Figure~3. Since the observations
were often made under different atmospheric conditions they also
provide a means of assessing the degree to which any contamination by
nuclear light can be attributed to either the airmass and seeing
conditions at the time of the observations or scattering within the
host galaxy itself. Where a different slit position was used, the two
spectra allow us to examine the degree of homogeneity in the stellar
composition of the galaxy.

It should be noted that due to the optimization method used to extract
the WHT spectra, their flux calibration can only be considered as
relative, not absolute. Different slit widths were also used on the
two instruments.  Comparisons of the flux densities obtained on the
two telescopes are therefore not meaningful.

\noindent
{\bf 0054$+$144 (RQQ):} different slit positions were used for the
Mayall 4-m and WHT observations. The spectrum obtained with the Mayall
4-m shows a contribution from H$\beta$/[O{\sc iii}] at $\lambda_{obs}
\sim5800$\AA, which is lacking in the WHT data. However, the
underlying continuum is very similar in both spectra and the measured
depth of the (weak) 4000\AA~break is very similar in each. This
implies that the emission lines detected in the Mayall 4-m spectrum
are a local feature, produced {\it in situ} rather than being due to
scattered light from emission-line regions in the nucleus, since they
do not appear to be accompanied by a corresponding amount of scattered
nuclear continuum. The consistency in the strength of the
4000\AA~break at the two observing epochs, along with its relative
weakness, suggests the presence of a significant population of young
stars in the host galaxy. McLure et al. (1999) report a tidal feature
visible in their $R$-band HST image of the elliptical host galaxy
which may be linked to the origin of this young stellar population.

\noindent
{\bf 0157$+$001 (RQQ):} the slit positions used in the two sets of
observations were essentially the same. However, the WHT spectrum
shows excess blue continuum shortwards of this feature. The two
spectra were taken under what were ostensibly very similar atmospheric
conditions, but we note that a very good approximation of the WHT
spectrum can be obtained simply by adding a scaled version of the
nuclear spectrum to the off-nuclear Mayall 4-m data. This suggests
that atmospheric scattering is to blame for the blue excess in the WHT
spectra.

\noindent
{\bf 0736$+$017 (RLQ):} the two spectra were both taken under good
atmospheric conditions and using the same slit position. The agreement
between the two is excellent.

\noindent
{\bf 2141$+$279 (RG):} the WHT spectrum has a much lower signal:noise
ratio than the Mayall 4-m spectrum, and uses a different slit
position. However, the overall continuum shape is in good agreement
with the earlier data. Both slits intercept an extension to the NE of
the nucleus which may be a tidal feature caused by interaction with a
northern companion.

\noindent
{\bf 2247$+$140 (RLQ):} the same slit position was used for both
spectra, although the seeing during the WHT run was quite poor ($\sim
1.6''$). The two datasets are consistent with one another, with a
relatively strong 4000\AA~break and little sign of an additional blue
continuum component from either young stars or scattered quasar
light. 

\noindent
{\bf 2344$+$184 (RQQ):} different slit positions were used on the two
different instruments, and the WHT spectrum shows an excess of blue
emission and a correspondingly weaker 4000\AA~break. The seeing
conditions were actually worse during the M4M observations, making
atmospheric scattering of quasar light an unlikely culprit. Moreover,
we note that the nuclear spectrum of this low-luminosity quasar is not
particularly blue and itself shows prominent stellar continuum
features (see Figure 2). More likely is that the WHT slit crossed a
region of the galaxy containing a large number of young stars. The
host of 2344$+$184 is a disc galaxy, with a central bulge and
prominent spiral arms (McLure et al. 1999). (Hutchings et al. (1989)
also suggest the presence of a bar.)


\section{Summary}

We describe and present optical spectra of 26 galaxies hosting
powerful nuclear activity. The sample contains carefully matched
subsamples of all three types of powerful AGN - radio-quiet and
radio-loud quasars, and FR{\sc ii} radio galaxies - enabling us to
investigate the relationship between the host galaxy and the radio
properties of the resident AGN and also to test unified models for
radio-loud objects.

The spectra were taken 5~arcsec {\it off-nucleus} and, via a careful
choice of slit position, aim to maximise the amount of galaxy light
entering the instrument whilst avoiding contamination from the active
nucleus. In the majority of cases this approach appears to have been
successful; the continuum is clearly overwhelmingly stellar in origin
and even when the presence of scattered nuclear emission is suspected,
features such as the 4000\AA~break are still clearly discernable.  In
almost all objects we detect at least some stellar signatures.

A second paper (Nolan et al. 2000) will discuss spectrophotometric
modelling of the galaxy spectra in order to determine the ages and
starformation histories of their constituent stellar populations.

\section*{Acknowledgments}

The authors would like to thank the staff at KPNO and the ING for
their assistance during the observations, and the anonymous referee
whose comments and suggestions resulted in several improvements in the
final paper. DHH and MJK acknowledge PPARC support. This research has
made use of the NASA/IPAC Extragalactic Database (NED), which is
operated by the Jet Propulsion Laboratory, California Institute of
Technology, under contract with the National Aeronautics and Space
Administration.

\section*{References}

\noindent
Bahcall J. N., Kirhakos S. \& Schneider D. P., 1994, ApJ, 435, L11
 
\noindent
Bahcall J. N., Kirhakos S. \& Schneider D. P., 1995a, ApJ, 447, L1
  
\noindent
Bahcall J. N., Kirhakos S. \& Schneider D. P., 1995b, ApJ, 450, 486
 
\noindent
Bahcall J. N., Kirhakos S. \& Schneider D. P., 1996, ApJ, 457, 557
 
\noindent
Bahcall J. N., Kirhakos S., Schneider D. P., \& Saxe D. H., 1997, ApJ, 479, 642
 
\noindent
Boroson T. A. \& Oke J. B., 1982, Nature, 296, 397

\noindent
Boroson T. A., Oke J. B. \& Green R. F., 1982, ApJ, 263, 32

\noindent
Boroson T. A. \& Oke J. B., 1984, ApJ, 281, 535

\noindent
Boroson T. A., Persson S. E. \& Oke J. B., 1985, ApJ, 293, 120

\noindent
Boyce P. J. et al., 1998, MNRAS, 298, 121

\noindent
Bruzual G., 1983, ApJ, 273, 105

\noindent
Disney M. J. et al., 1995, Nature, 376, 150

\noindent
Dunlop J. S., McLure R. J., Kukula M. J., Baum S. A., O'Dea C. P. \& Hughes D. H., 2000, in preparation	

\noindent
Dunlop J. S., Taylor G. L., Hughes D. H. \& Robson E. I., 1993, MNRAS, 264, 455

\noindent
Green R. F., Williams T. B. \& Morton D. C., 1978, ApJ, 226, 729

\noindent
Hamilton D., 1985, ApJ, 297, 371

\noindent
Heckman T. M., Bothun G. D., Balick B. \& Smith E. P., 1984, AJ, 89, 958

\noindent
Hickson P. \& Hutchings J. B., 1987, ApJ, 312, 518

\noindent 
Hooper E. J., Impey C. D. \& Foltz C. B., 1997, ApJ, 480, L95

\noindent
Hutchings J. B., Crampton D., Campbell B., Duncan D. \& Glendenning B., 1984, ApJS, 55, 319

\noindent
Hutchings J. B. \& Crampton D., 1990, AJ, 99, 37
 
\noindent
Hutchings J. B. et al., 1994, ApJ, 429, L1

\noindent
Hutchings J. B., Janson T. \& Neff S. G., 1989, ApJ, 342, 660

\noindent 
Hutchings J. B. \& Morris S. C., 1995, AJ, 109, 1541

\noindent
Kukula M. J., Dunlop J. S., Hughes D. H., Taylor G. L. \& Boroson T.,
1997, in `Quasar Hosts', eds. Clements D. L. \& P\'{e}rez-Fournon I.,
Springer:Berlin, p. 177

\noindent
Kukula M. J., Dunlop J. S., McLure R. J., Baum S. A., O'Dea C. P. \&
Hughes D. H., 1999, Advances In Space Research, 23, 1131

\noindent
MacKenty J. W. \& Stockton A., 1984, ApJ, 283, 64


\noindent
McHardy I. M., Merrifield M. R., Abraham R. G. \& Crawford C. S., 1994, MNRAS, 268, 681

\noindent 
McLeod K. K. \& Rieke G. H., 1995a, ApJ, 441, 96

\noindent
McLeod K. K. \& Rieke G. H., 1995b, ApJ, 454, L77

\noindent
McLeod K. K.,  Rieke G. H. \& Storrie-Lombardi L., 1999, ApJ, 511, L67

\noindent
McLure R. J., Kukula M. J., Dunlop J. S., Baum S. A., O'Dea C. P. \& Hughes D. H., 1999, MNRAS, 308, 377

\noindent
Meurs E. J. A. \& Unger S. W., 1991, A\&A, 252, 63

\noindent
Nolan L. A., Dunlop J. S., Kukula M. J., Hughes D. H., Boroson T., \& Jimenez R., 2000, submitted to MNRAS

\noindent
Stockton A., 1982, ApJ, 257, 33

\noindent
Stockton A. \& Farnham T., 1991, ApJ, 371, 525

\noindent
Stockton A. \& MacKenty J. W., 1987, ApJ, 316, 584

\noindent
Taylor G. L., Dunlop J. S., Hughes D. H. \& Robson E. I., 1996, MNRAS, 283, 930

\noindent
Urry C. M. \& Padovani P., PASP, 107, 803

\noindent
V\'{e}ron-Cetty M. -P. \& Woltjer L., 1990, A\&A, 236, 69

\clearpage


\begin{figure*}
\vspace{21.0cm}
\includegraphics{figures/spectra/spec0007_m4m.ps}
\includegraphics{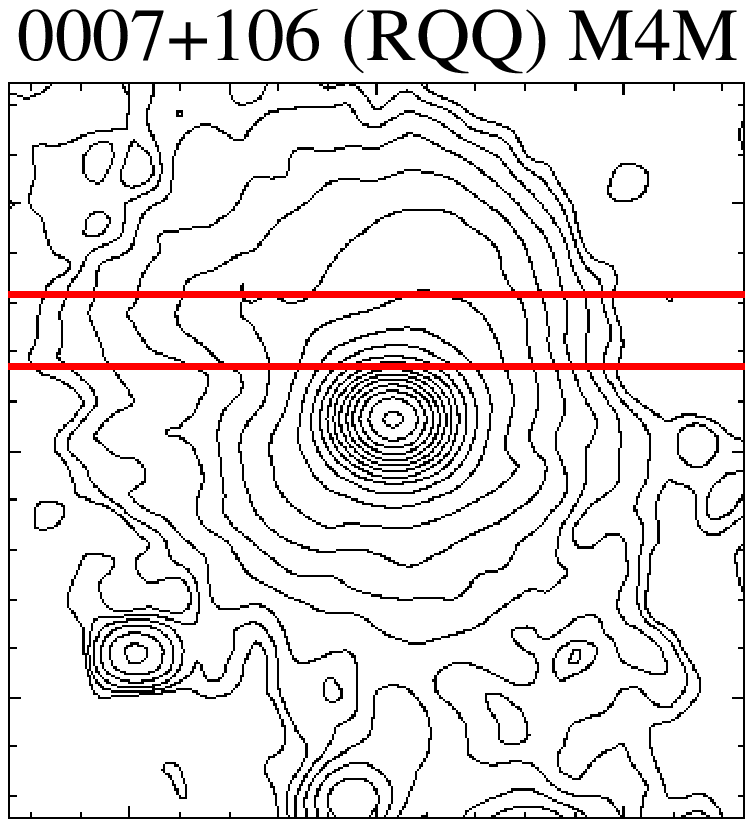}

\includegraphics{figures/spectra/spec0054_m4m.ps}
\includegraphics{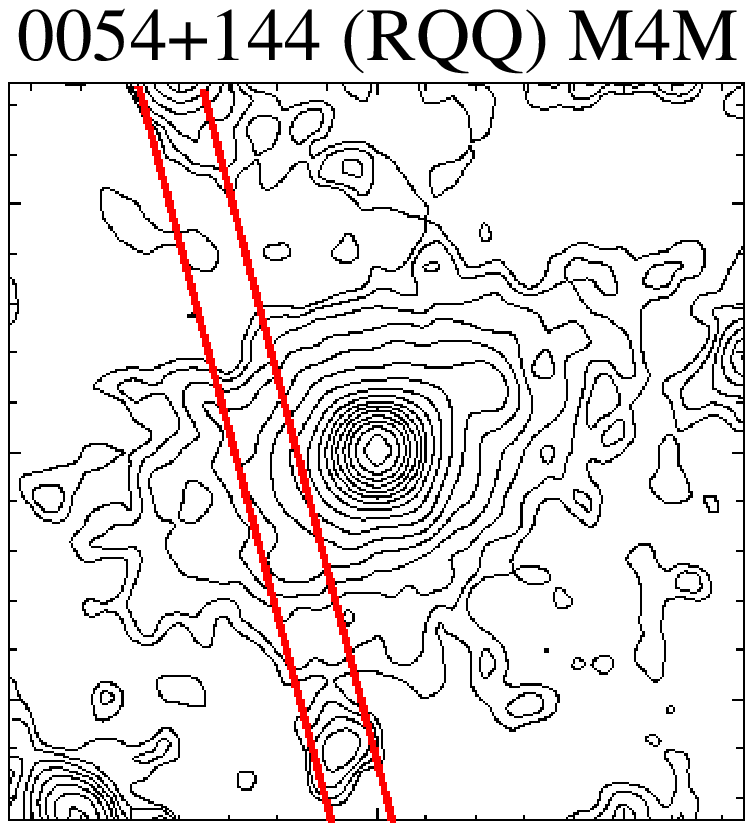}

\includegraphics{figures/spectra/spec0054_wht.ps}
\includegraphics{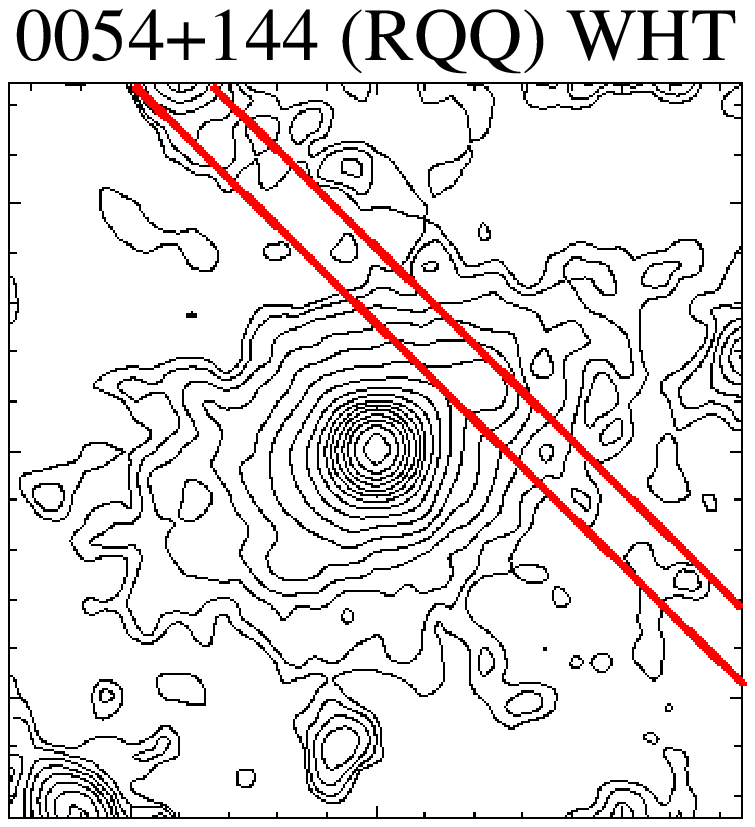}

\caption{Spectra obtained on the Mayall 4-m Telescope (M4M) at Kitt
Peak and the Willliam Herschel Telescope (WHT) on La Palma, in order
of increasing right ascension.  The spectra are plotted at the
observed wavelength in units of $F_{\lambda}$. In each case the lower
panel shows the off-nuclear (host galaxy) spectrum and the upper panel
the nuclear (quasar) spectrum where available. The expected positions
of various stellar absorption features are marked by vertical dotted
lines in the off-nuclear spectra: (a) the 4000\AA~break; (b) G band
($\lambda_{rest} = 4300 \rightarrow 4320$\AA); (c) Mg Ib
($\lambda_{rest} = 5173$\AA); (d) Fe$\lambda5270$.  The redshifted
wavelengths of the [O{\sc iii}]$\lambda5007$ and H$\alpha$ emission
lines are also indicated. The side panels show the slit position and
orientation for the off-nuclear data superimposed on the near-infrared
(2.2$\mu$m) contours of the object from Taylor et al. (1996) (each
panel is $30\times30$~arcsec$^{2}$).}

\end{figure*}

\begin{figure*}
\setcounter{figure}{1}
\vspace{21.0cm}
\includegraphics{figures/spectra/spec0137_m4m.ps}
\includegraphics{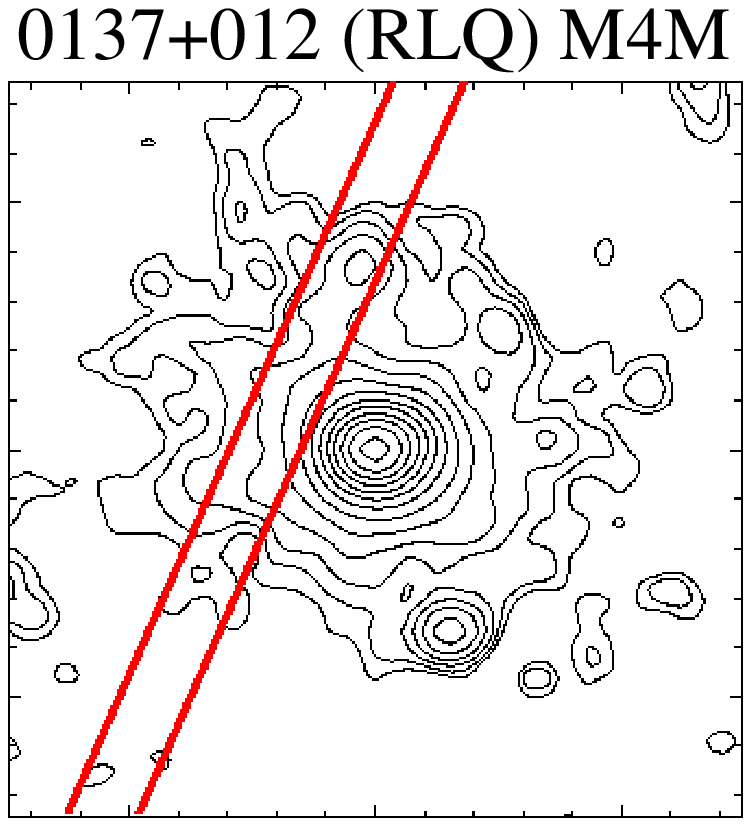}

\includegraphics{figures/spectra/spec0157_m4m.ps}
\includegraphics{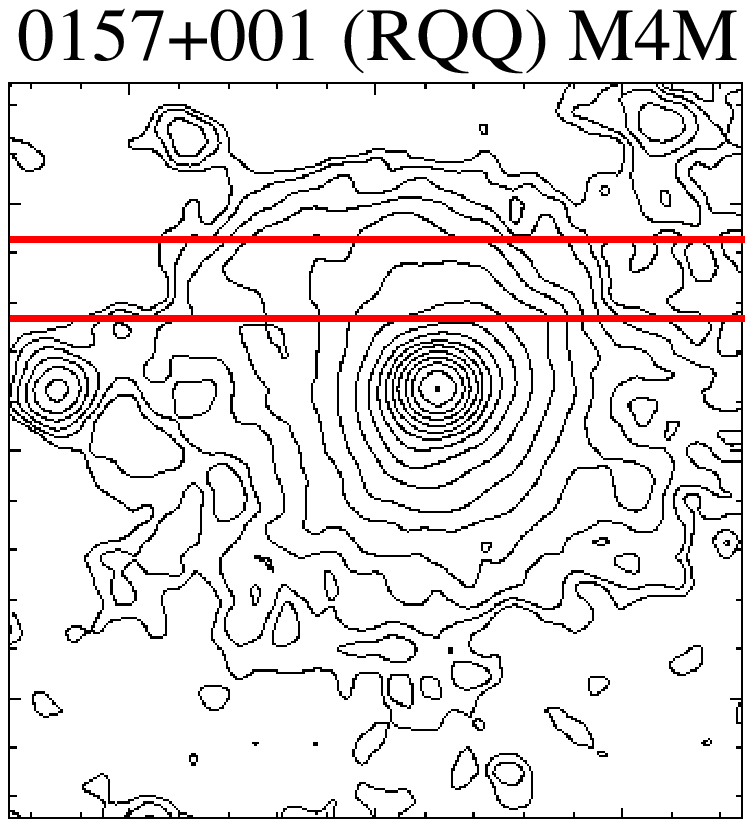}

\includegraphics{figures/spectra/spec0157_wht.ps}
\includegraphics{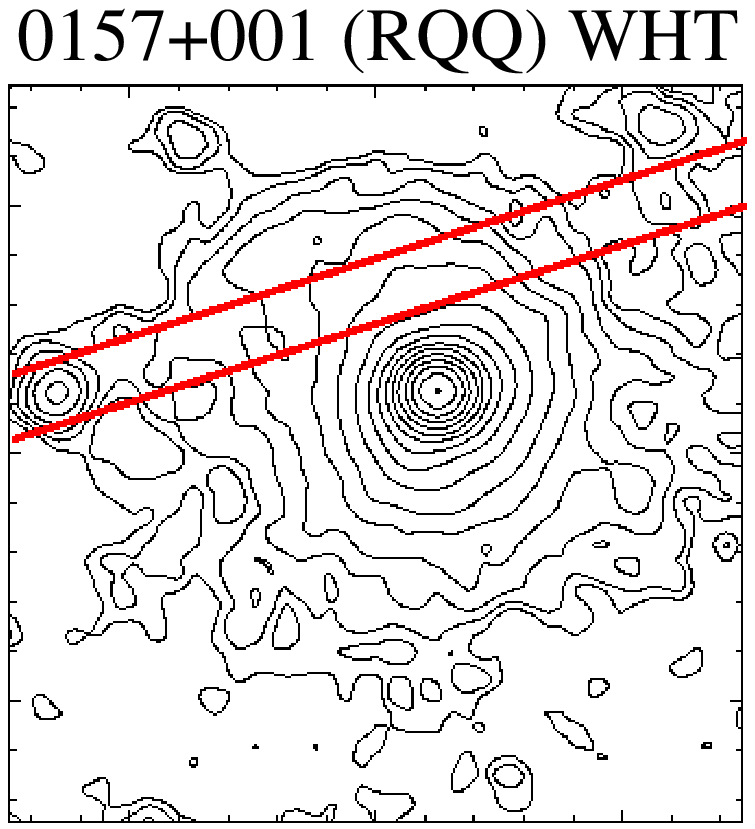}

\caption{continued.}

\end{figure*}

\begin{figure*}
\setcounter{figure}{1}
\vspace{21.0cm}
\includegraphics{figures/spectra/spec0204_wht.ps}
\includegraphics{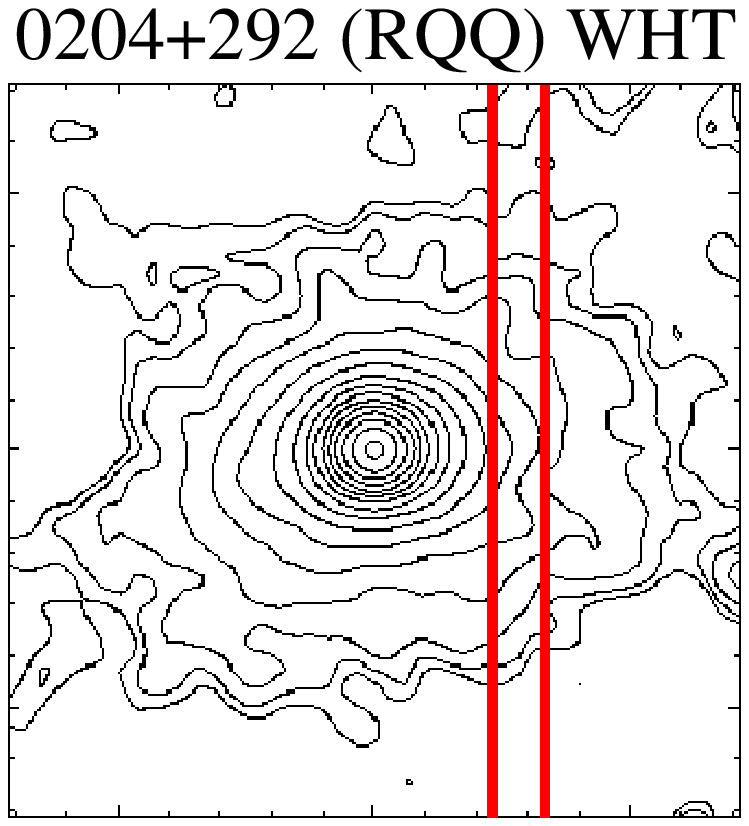}

\includegraphics{figures/spectra/spec0230_wht.ps}
\includegraphics{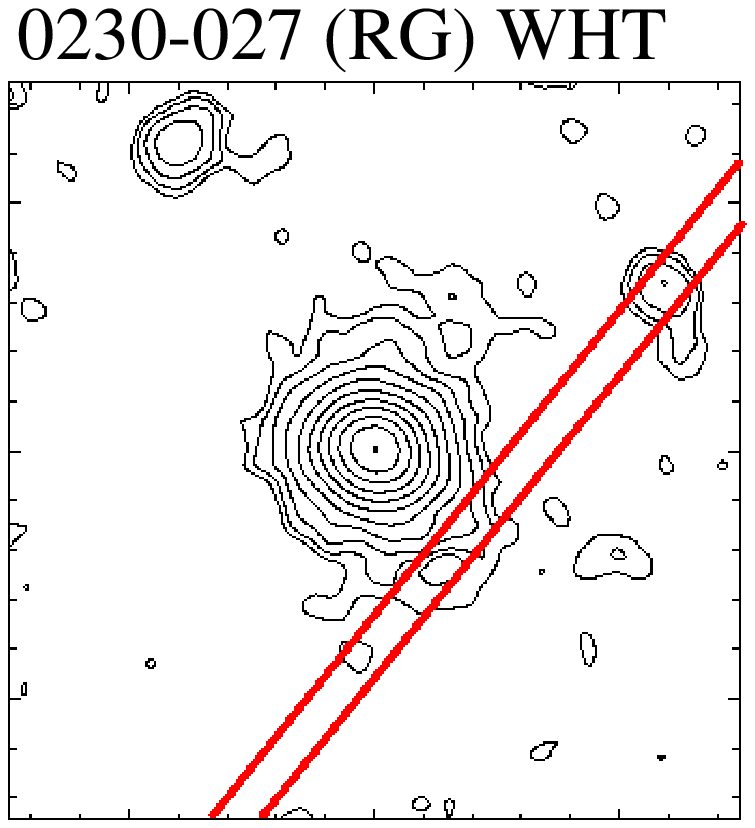}

\includegraphics{figures/spectra/spec0244_wht.ps}
\includegraphics{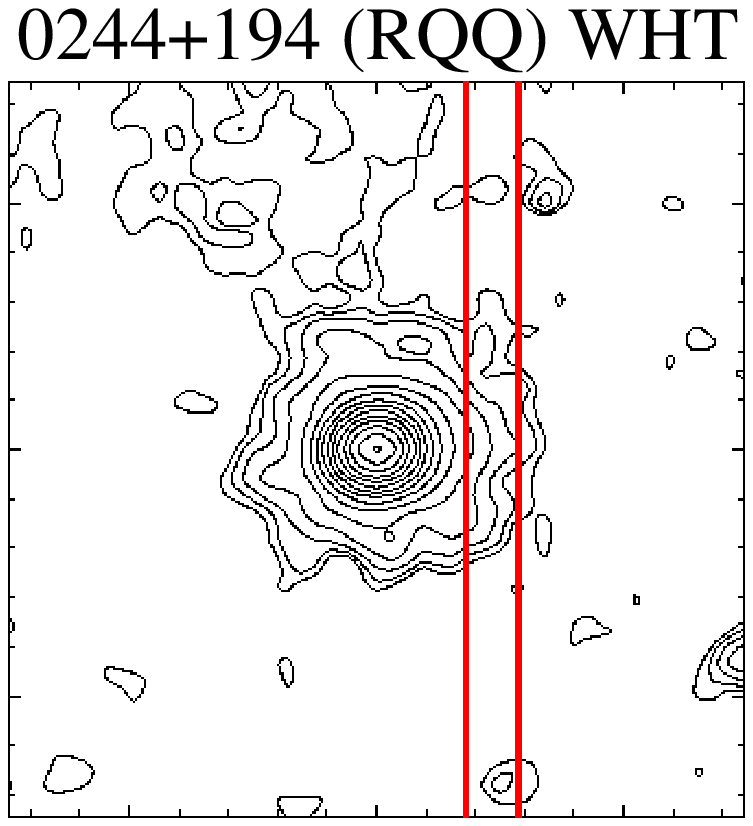}

\caption{continued.}

\end{figure*}

\begin{figure*}
\setcounter{figure}{1}
\vspace{21.0cm}
\includegraphics{figures/spectra/spec0345_wht.ps}
\includegraphics{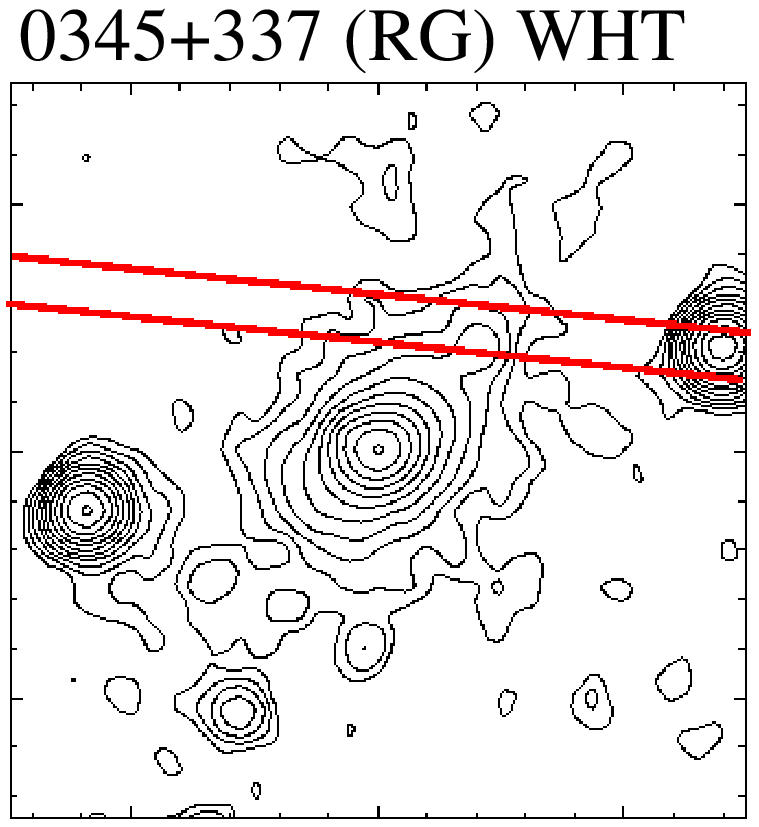}

\includegraphics{figures/spectra/spec0736_m4m.ps}
\includegraphics{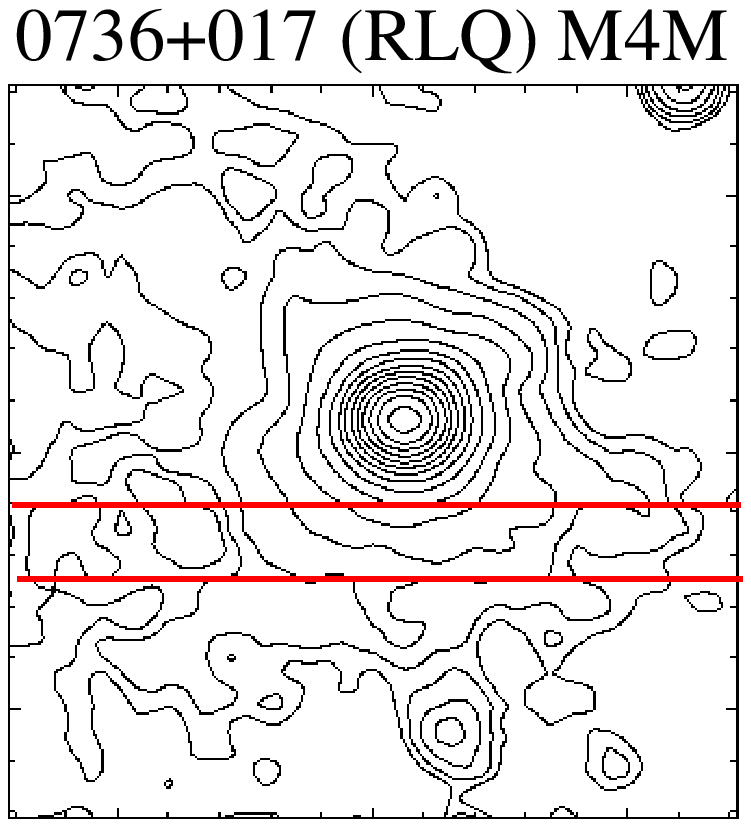}

\includegraphics{figures/spectra/spec0736_wht.ps}
\includegraphics{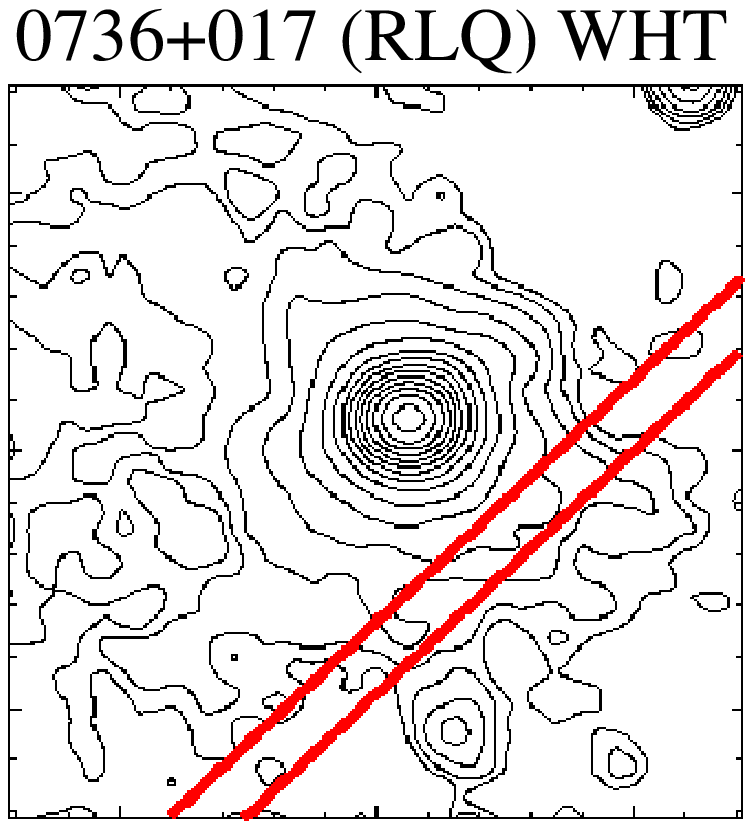}

\caption{continued.}

\end{figure*}

\begin{figure*}
\setcounter{figure}{1}
\vspace{21.0cm}
\includegraphics{figures/spectra/spec0917_wht.ps}
\includegraphics{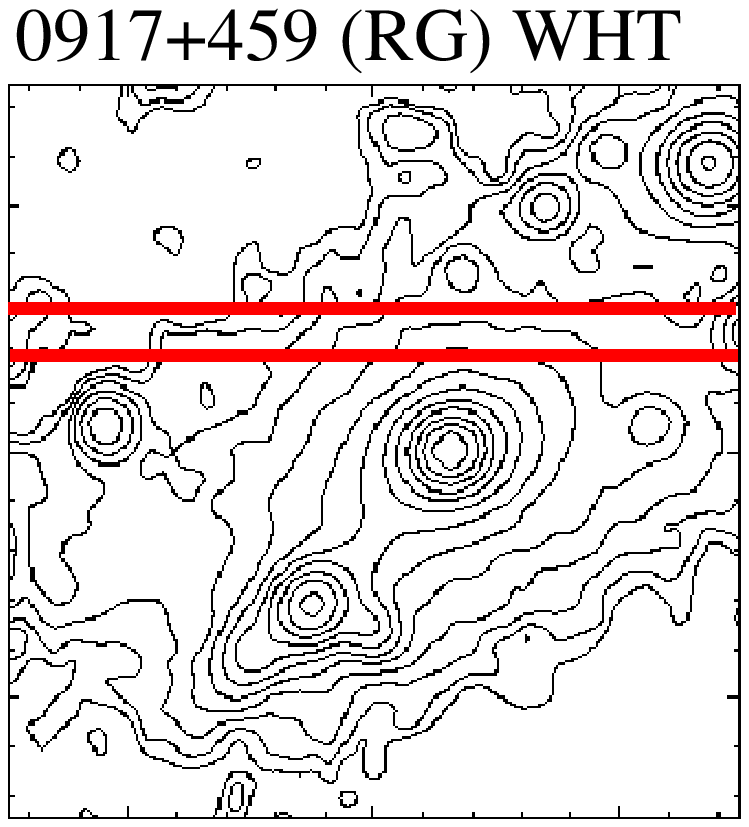}

\includegraphics{figures/spectra/spec1004_wht.ps}
\includegraphics{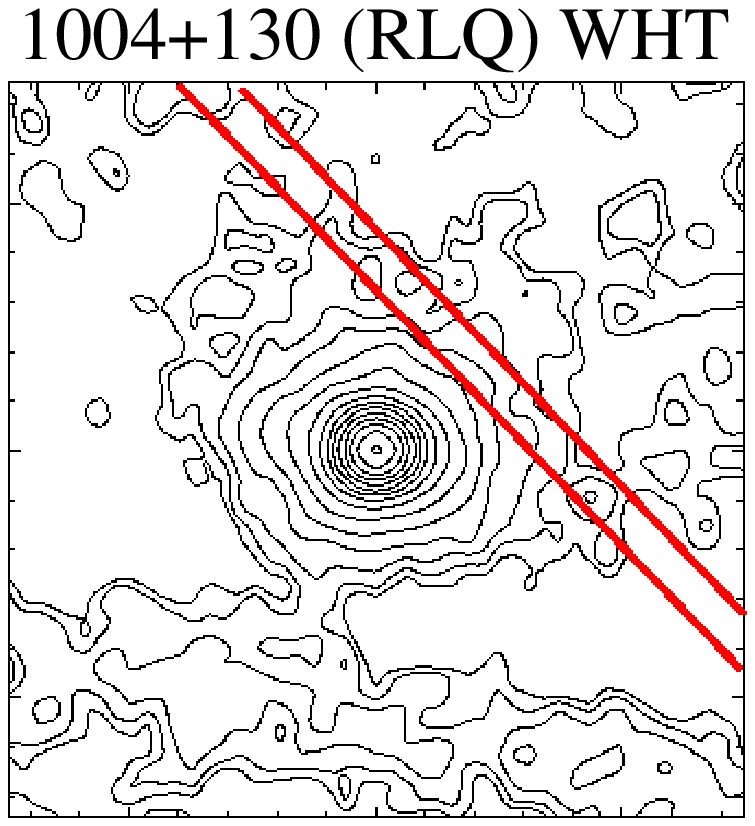}

\includegraphics{figures/spectra/spec1020_m4m.ps}
\includegraphics{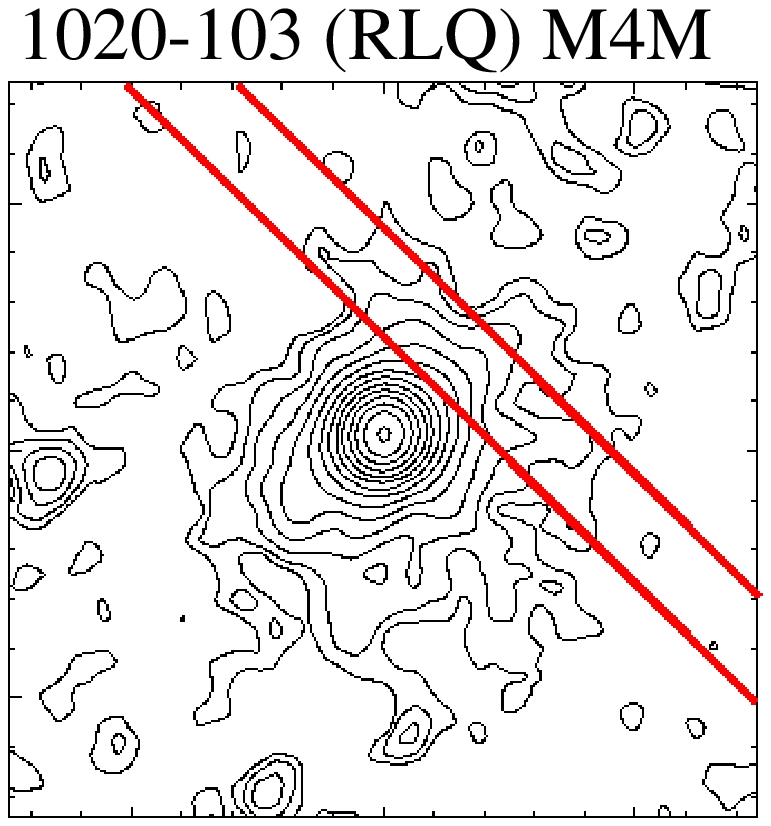}

\caption{continued.}

\end{figure*}

\begin{figure*}
\setcounter{figure}{1}
\vspace{21.0cm}
\includegraphics{figures/spectra/spec1215_wht.ps}
\includegraphics{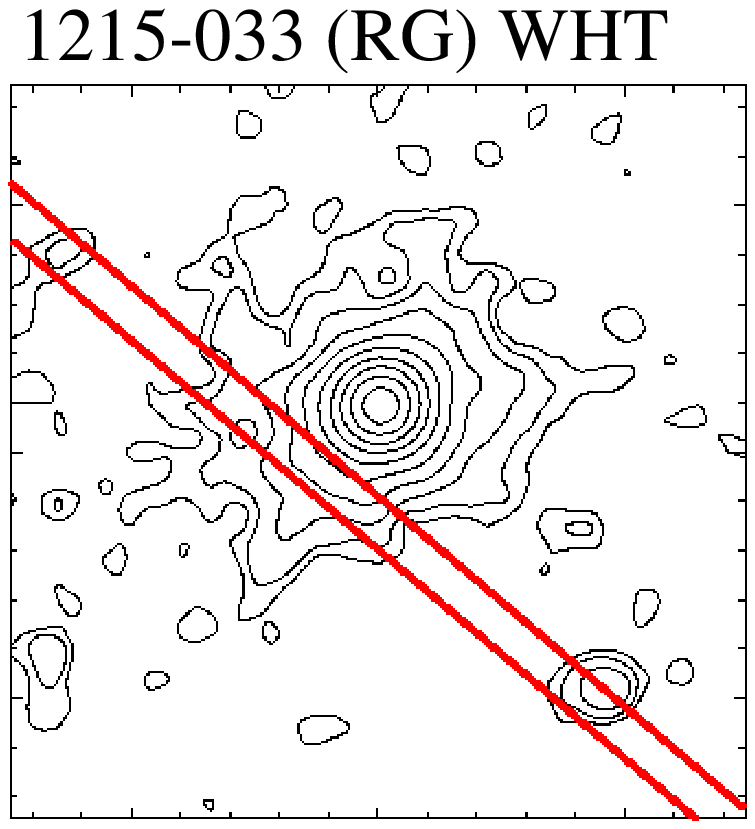}

\includegraphics{figures/spectra/spec1217_wht.ps}
\includegraphics{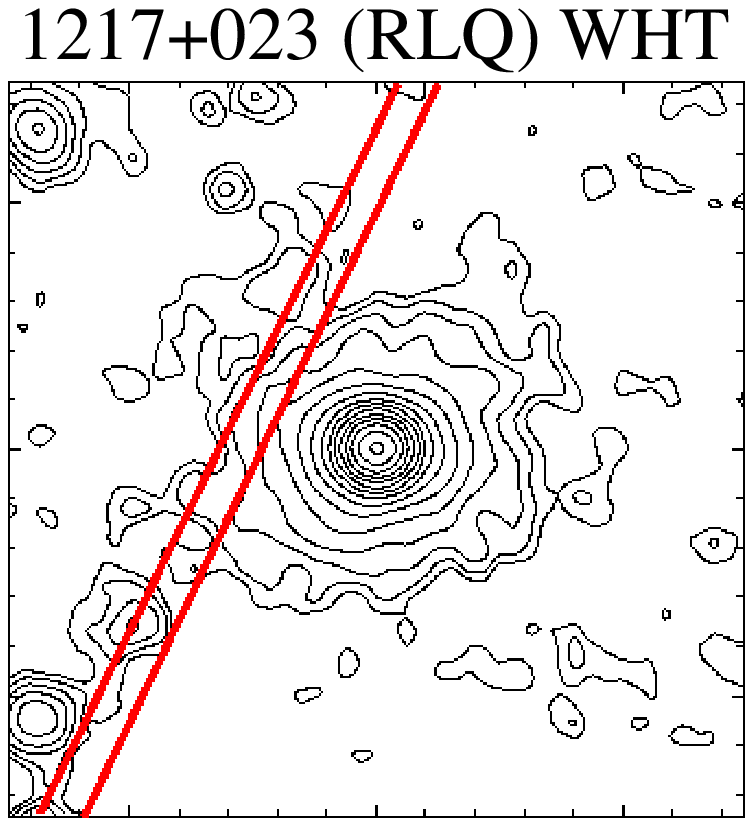}

\includegraphics{figures/spectra/spec1330_m4m.ps}
\includegraphics{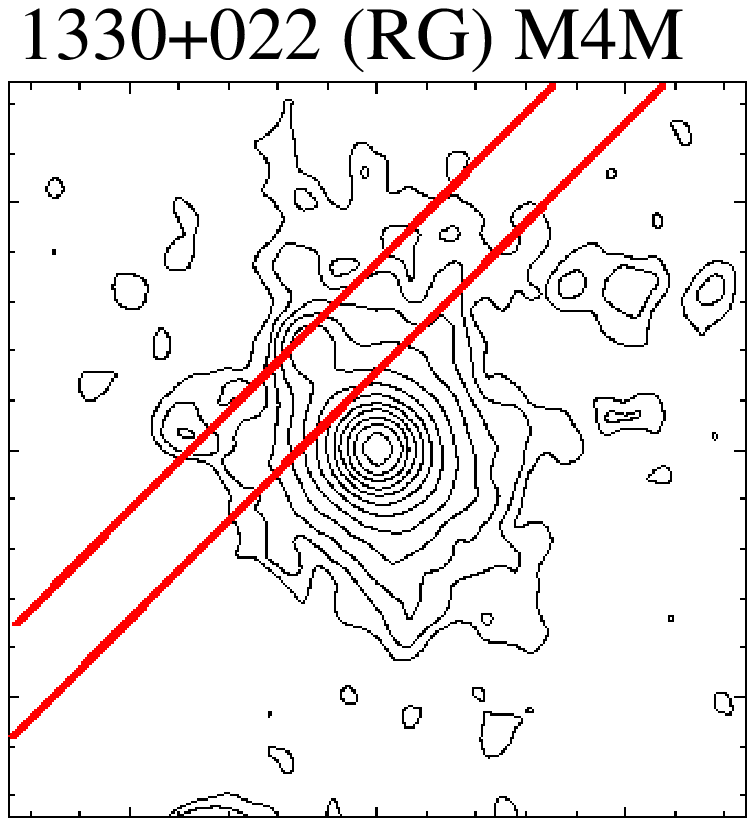}

\caption{continued.}

\end{figure*}

\begin{figure*}
\setcounter{figure}{1}
\vspace{21.0cm}
\includegraphics{figures/spectra/spec1334_wht.ps}
\includegraphics{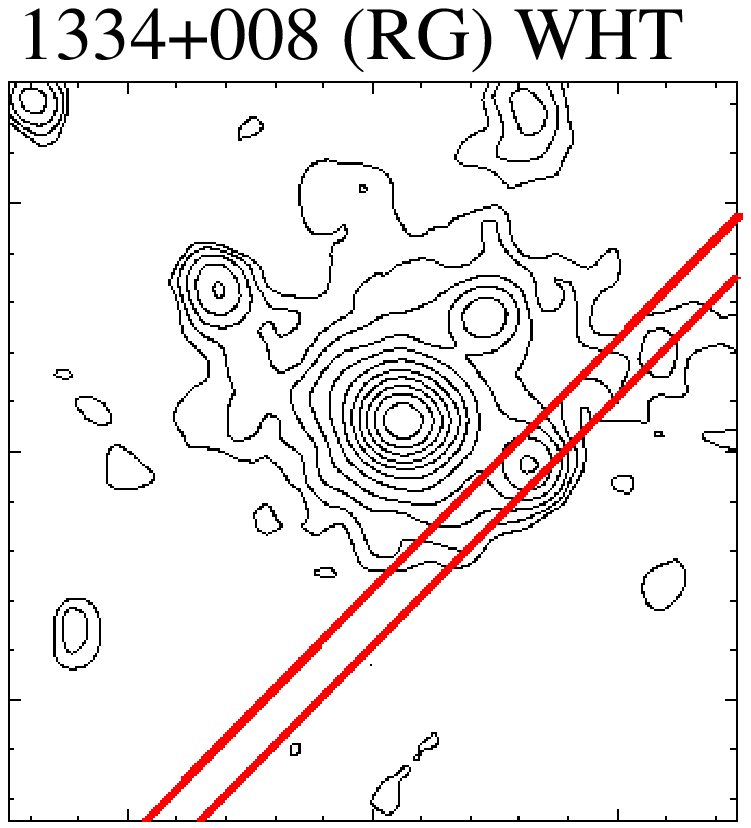}

\includegraphics{figures/spectra/spec1549_wht.ps}
\includegraphics{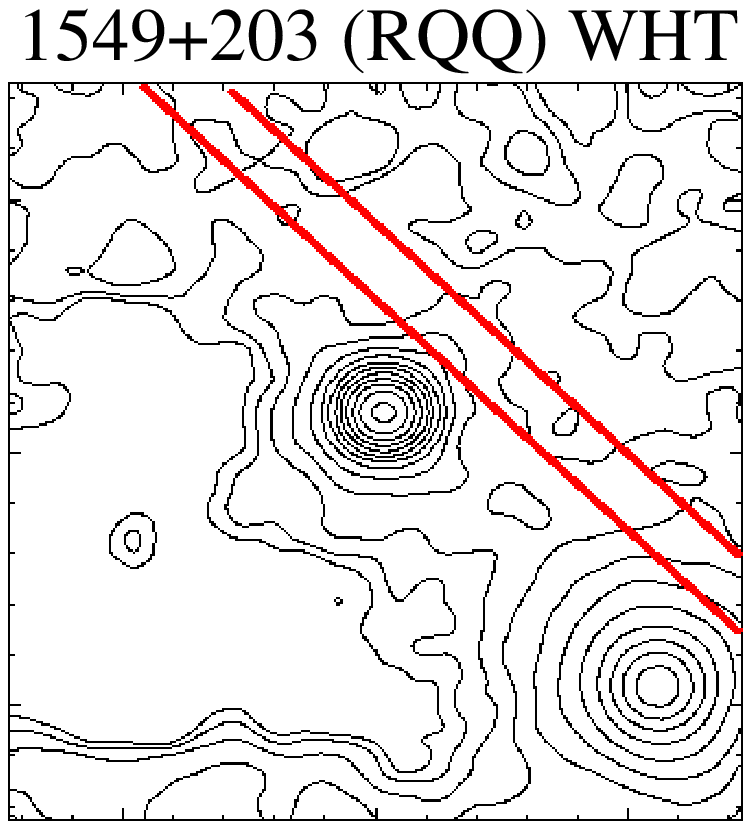}

\includegraphics{figures/spectra/spec1635_wht.ps}
\includegraphics{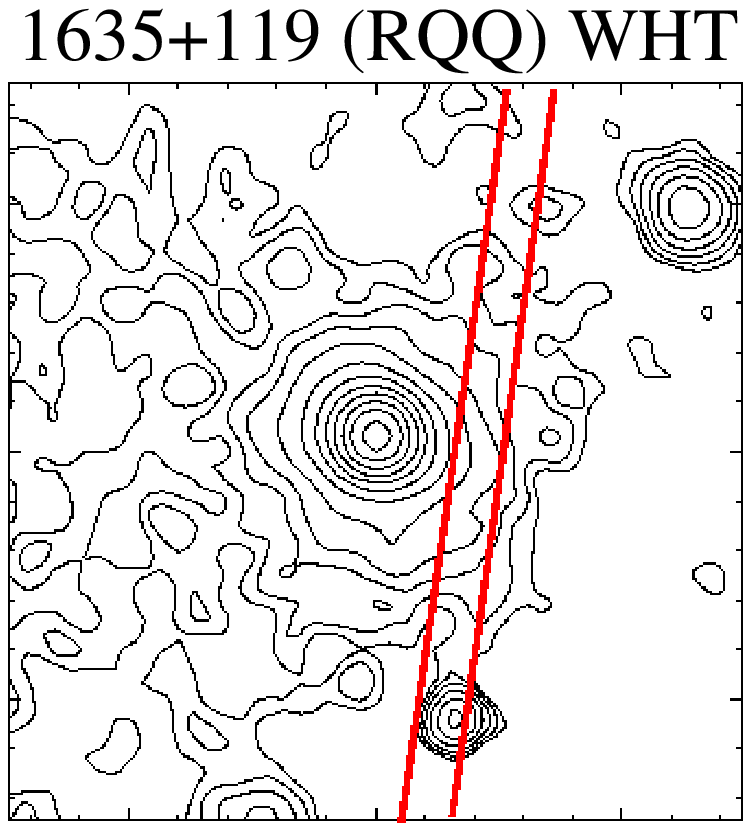}

\caption{continued.}

\end{figure*}

\begin{figure*}
\setcounter{figure}{1}
\vspace{21.0cm}
\includegraphics{figures/spectra/spec2135_wht.ps}
\includegraphics{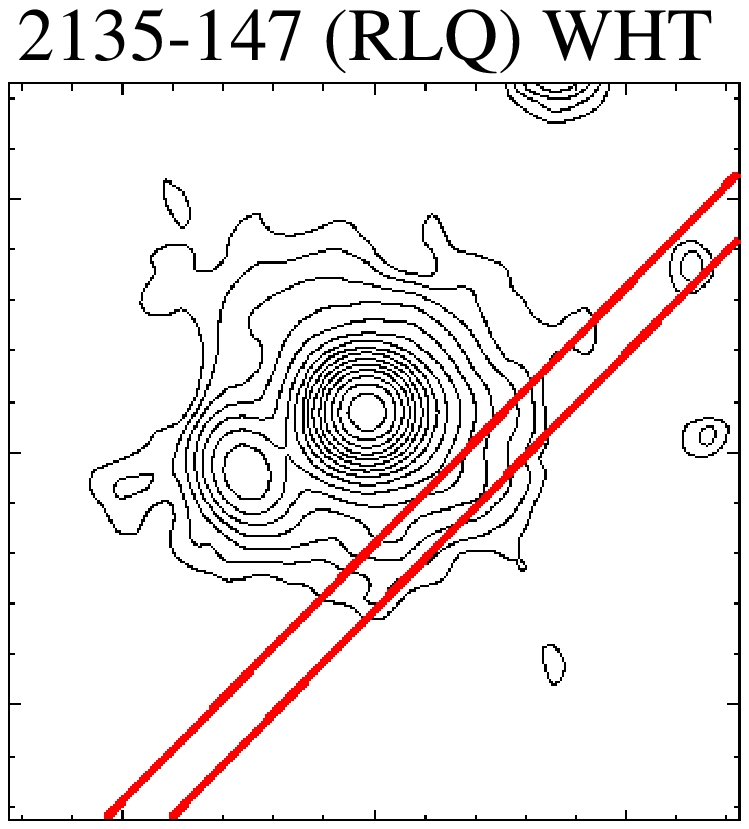}

\includegraphics{figures/spectra/spec2141+175_wht.ps}
\includegraphics{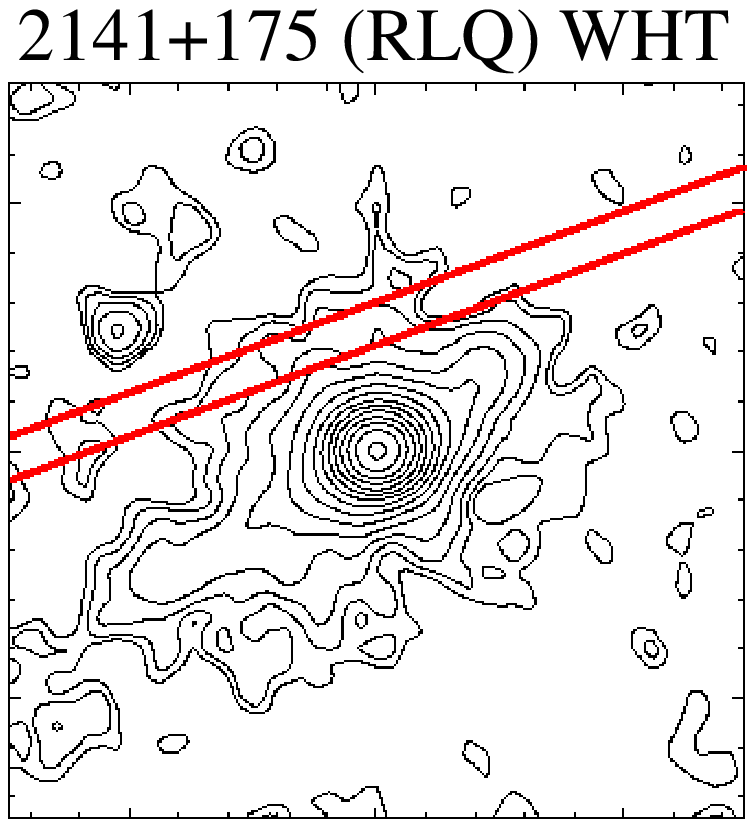}

\includegraphics{figures/spectra/spec2141+279_m4m.ps}
\includegraphics{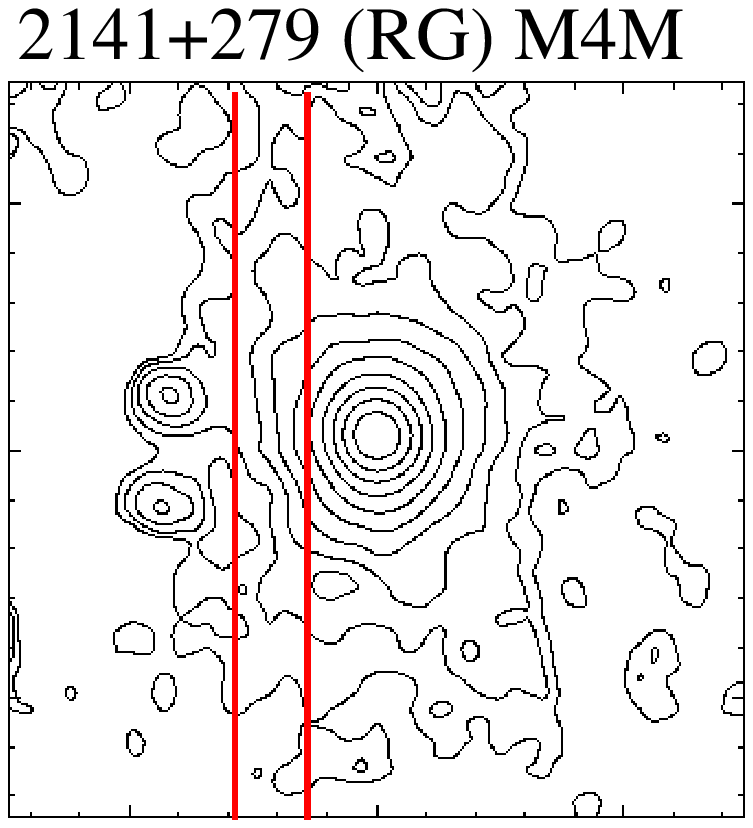}

\caption{continued.}

\end{figure*}

\begin{figure*}
\setcounter{figure}{1}
\vspace{21.0cm}
\includegraphics{figures/spectra/spec2141+279_wht.ps}
\includegraphics{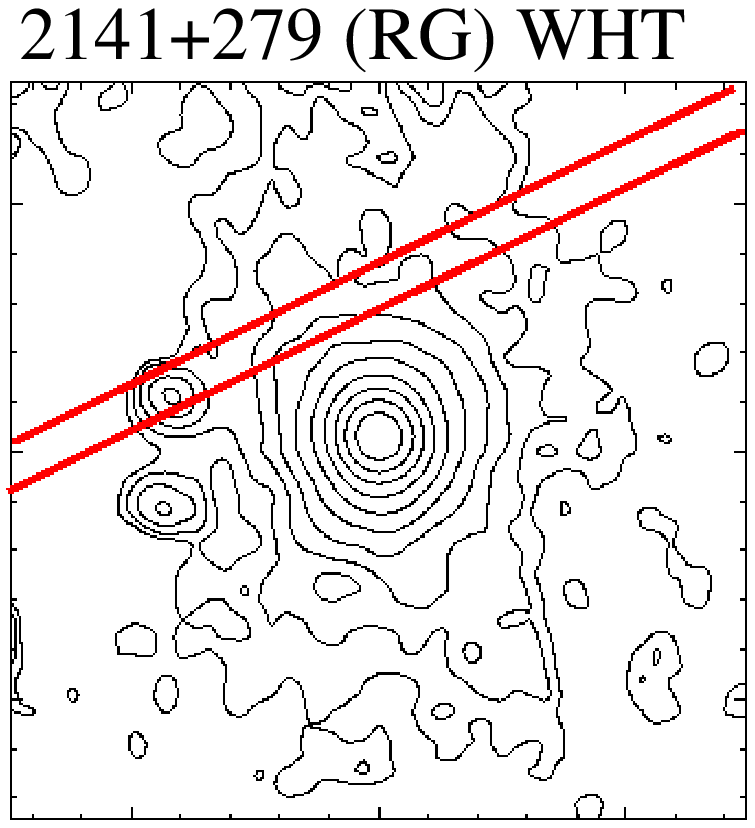}

\includegraphics{figures/spectra/spec2201_m4m.ps}
\includegraphics{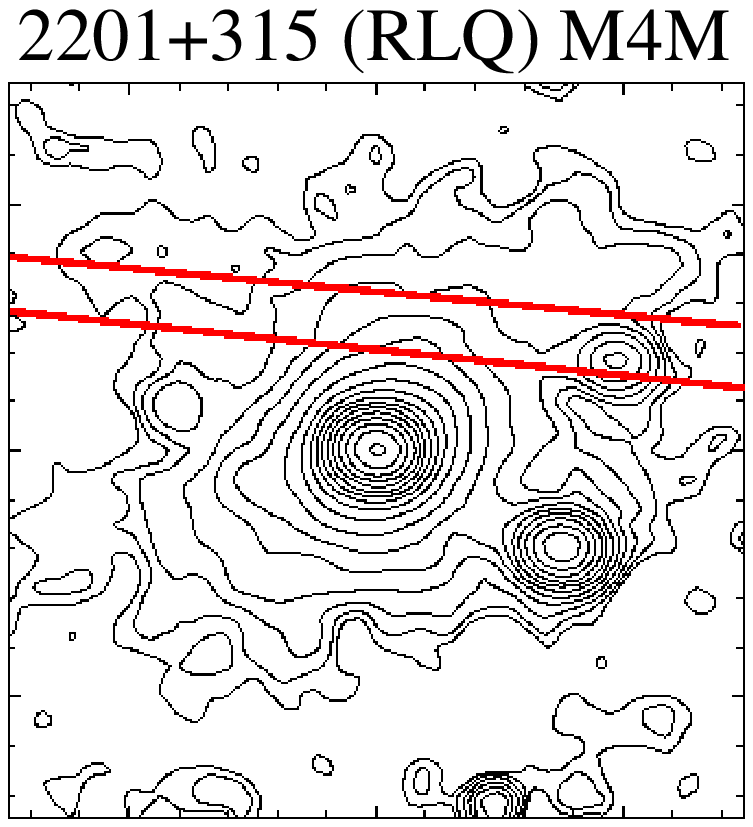}

\includegraphics{figures/spectra/spec2215_wht.ps}
\includegraphics{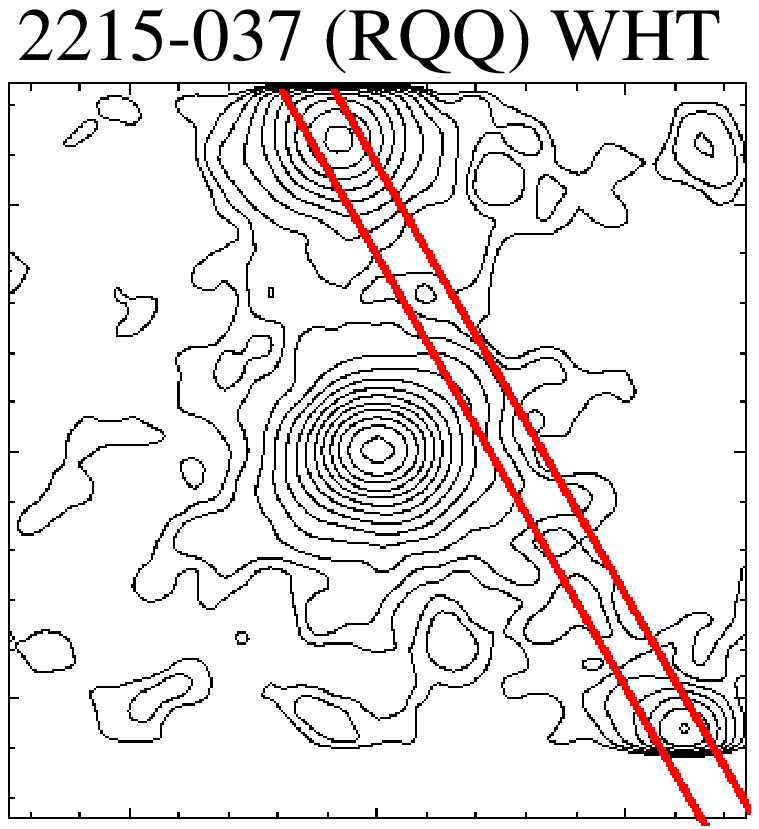}

\caption{continued.}

\end{figure*}

\begin{figure*}
\setcounter{figure}{1}
\vspace{21.0cm}
\includegraphics{figures/spectra/spec2247_m4m.ps}
\includegraphics{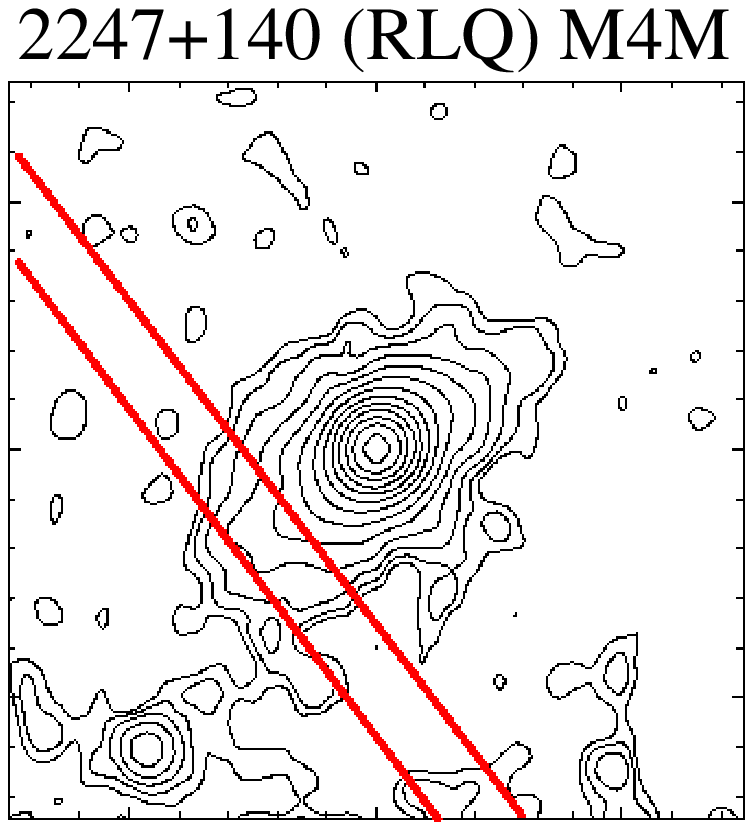}

\includegraphics{figures/spectra/spec2247_wht.ps}
\includegraphics{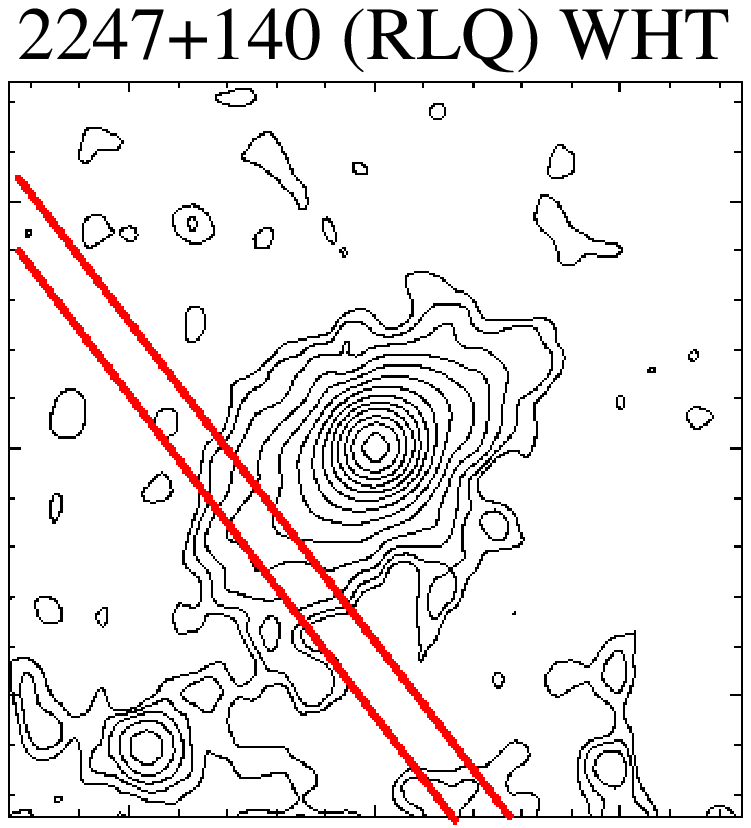}

\includegraphics{figures/spectra/spec2344_m4m.ps}
\includegraphics{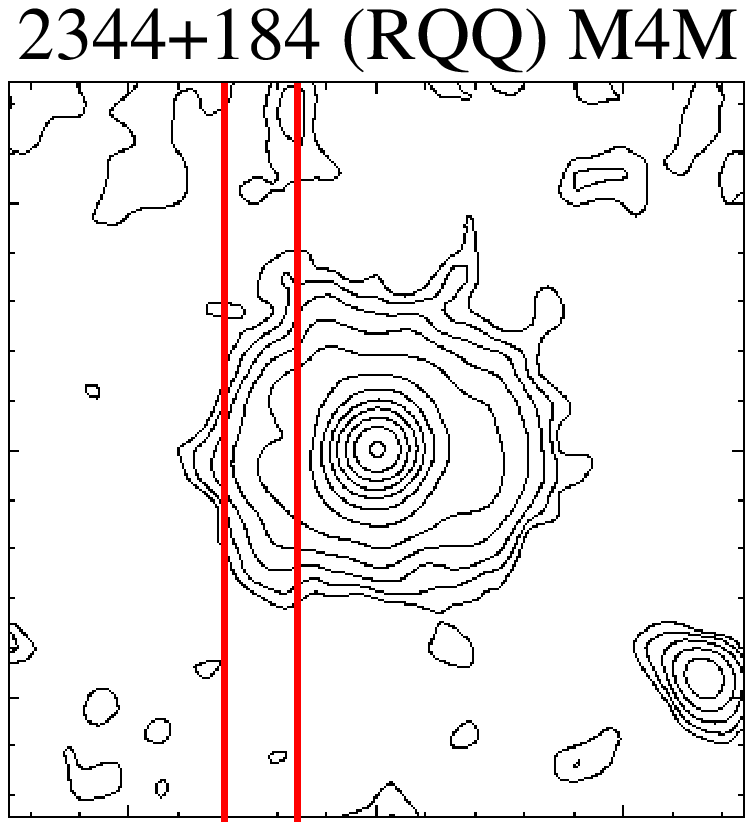}

\caption{continued.}

\end{figure*}

\begin{figure*}
\setcounter{figure}{1}
\vspace{14.0cm}
\includegraphics{figures/spectra/spec2344_wht.ps}
\includegraphics{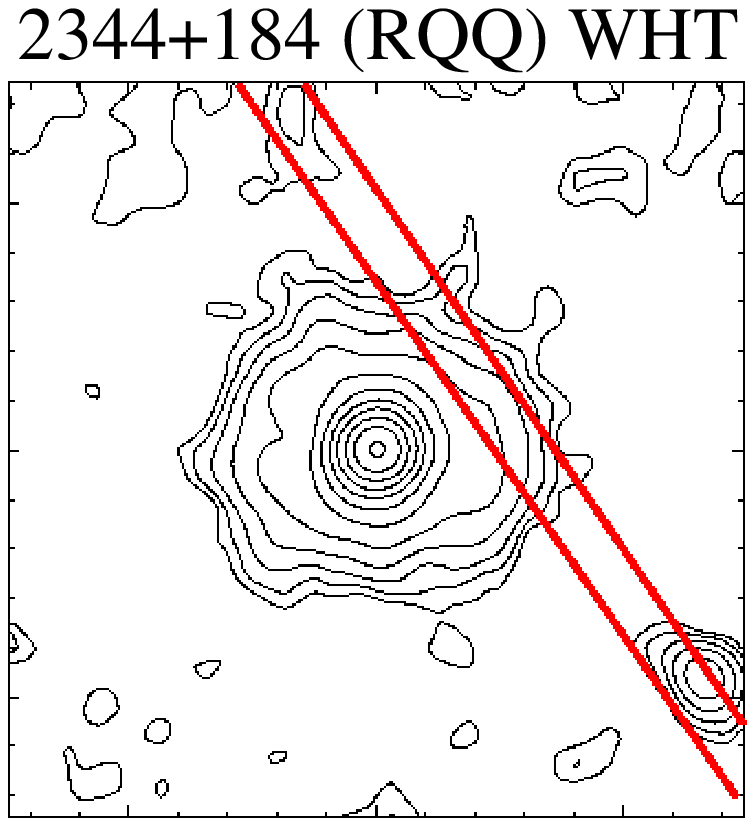}

\includegraphics{figures/spectra/spec2349_wht.ps}
\includegraphics{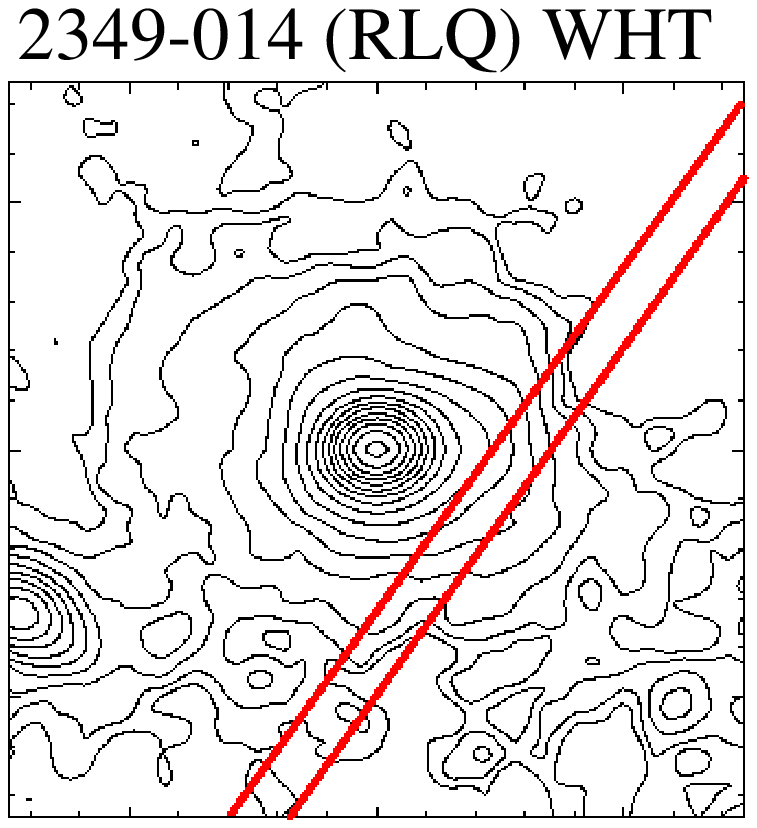}

\caption{continued.}

\end{figure*}

\begin{figure*}
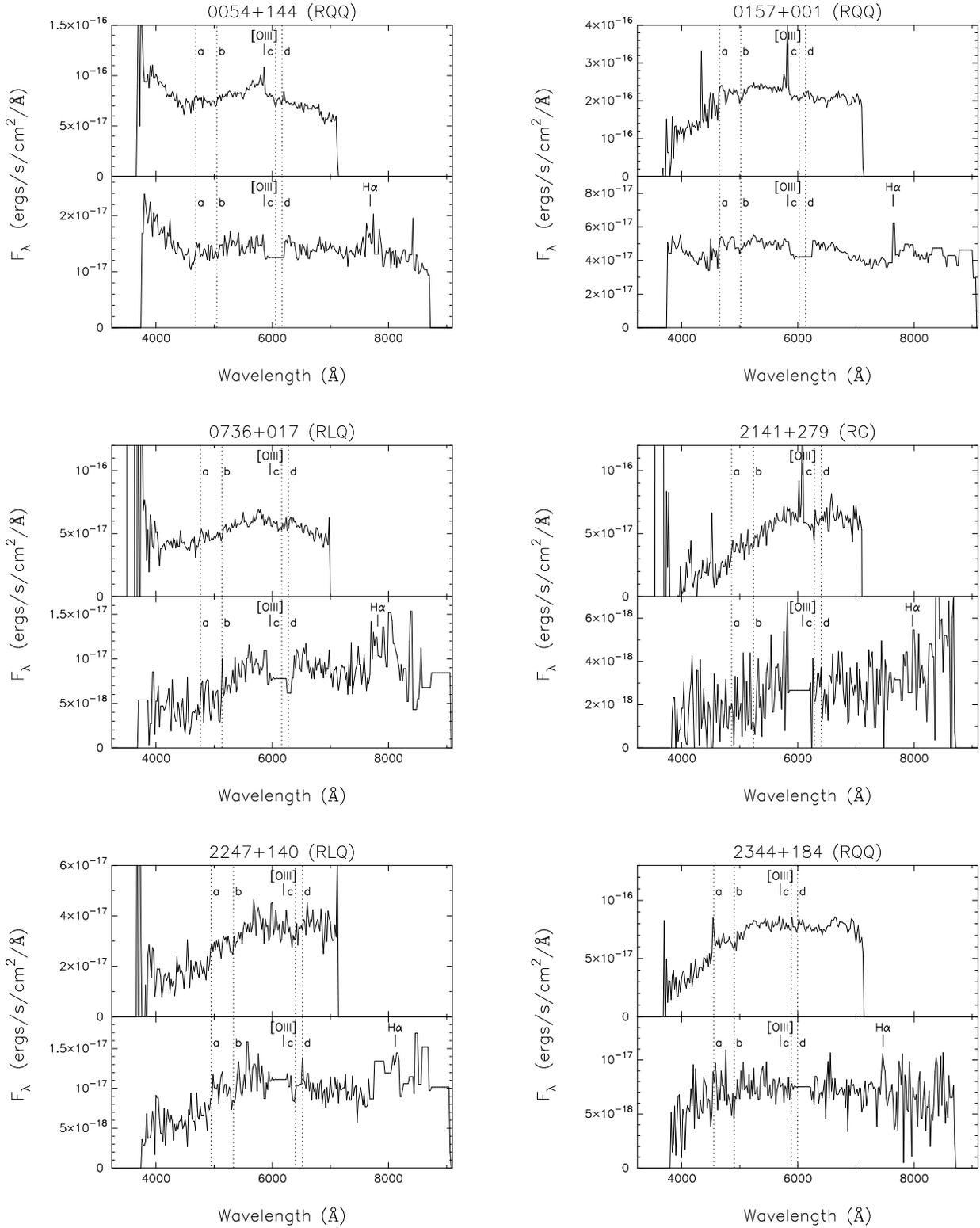

\setcounter{figure}{2}
\vspace{21.0cm}
\includegraphics{figures/compare/spec0054_m4m_wht.ps}
\includegraphics{figures/compare/spec0157_m4m_wht.ps}

\includegraphics{figures/compare/spec0736_m4m_wht.ps}
\includegraphics{figures/compare/spec2141+279_m4m_wht.ps}

\includegraphics{figures/compare/spec2247_m4m_wht.ps}
\includegraphics{figures/compare/spec2344_m4m_wht.ps}

\caption{A comparison of off-nuclear (host galaxy) spectra obtained on
the Mayall 4-m Telescope at Kitt Peak (upper panel) with those
obtained with the 4.2m WHT (lower panel). The spectra are plotted at
the observed wavelength in units of $F_{\lambda}$. The expected
positions of various stellar absoption features are shown as vertical
dotted lines: (a) the 4000\AA~break; (b) G band ($\lambda_{rest} =
4300 \rightarrow 4320$\AA); (c) Mg Ib ($\lambda_{rest} = 5173$\AA);
(d) Fe$\lambda5270$.  The redshifted wavelengths of the [O{\sc
iii}]$\lambda5007$ and H$\alpha$ emission lines are also
marked. Different slit widths were used on the two instruments and the
flux callibration for the WHT spectra is relative rather than
absolute, so the flux values from the two telescopes are not directly
comparable.}

\end{figure*}

\end{document}